\documentclass[superscriptaddress, showkeys, prc]{revtex4-2}

\usepackage{eurosym}
\usepackage{hyperref}
\hypersetup{
	unicode=true, 
	colorlinks=true, 
	linkcolor=blue, 
	citecolor=blue, 
	urlcolor=blue 
}
\usepackage{amsmath,amsfonts,amssymb,amsthm,mathtools}
\usepackage{txfonts}
\usepackage{icomma}
\usepackage{fdsymbol}
\usepackage[dvipsnames]{xcolor}
\usepackage{graphicx}
\usepackage{dcolumn}
\usepackage{bm}
\usepackage{wrapfig}
\usepackage{array,tabularx,tabulary,booktabs}
\usepackage{longtable}
\usepackage{multirow}
\usepackage[normalem]{ulem}

\graphicspath{{images/}}
\newcolumntype{M}[1]{>{\centering\arraybackslash}m{#1}}

\begin{document}
		
		\title{The astrophysical $S-$factor and reaction rate for $^{15}$N($p,\gamma
			$)$^{16}$O within the modified potential cluster model}
		
		\author{S. B. Dubovichenko}
		\affiliation{Fesenkov Astrophysical Institute$,$ 050020 Almaty$,$ Kazakhstan}
		\author{A.S. Tkachenko}
		\affiliation{Fesenkov Astrophysical Institute$,$ 050020 Almaty$,$ Kazakhstan}
		\author{R. Ya. Kezerashvili}
		\affiliation{New York City College of Technology$,$ City University of New York$,$ Brooklyn$,$ 11201 New York$,$ USA}
		\affiliation{Graduate School and University Center$,$ City University of New York$,$ 10016 New York$,$ USA}
		\author{N.A. Burkova}
		\affiliation{al-Farabi Kazakh National University$,$ 050040 Almaty$,$ Kazakhstan}
		\author{B. M. Yeleusheva}
		\affiliation{Fesenkov Astrophysical Institute$,$ 050020 Almaty$,$ Kazakhstan}
		\affiliation{al-Farabi Kazakh National University$,$ 050040 Almaty$,$ Kazakhstan}
		
		\begin{abstract}
			We study a radiative $p^{15}$N capture on the ground state of $^{16}$O at
			stellar energies within the framework of a modified potential cluster model
			(MPCM) with forbidden states, including low-lying resonances. The
			investigation of the $^{15}$N($p,\gamma _{0}$)$^{16}$O reaction includes the
			consideration of $^{3}S_{1}$ resonances due to $E1$ transitions and the
			contribution of $^{3}P_{1}$ scattering wave in $p$ + $^{15}$N channel due to 
			$^{3}P_{1}\longrightarrow $ $^{3}P_{0}$ $M1$ transition. We calculated the
			astrophysical low-energy $S-$factor, and extrapolated $S(0)$ turned out to
			be within $34.7-40.4$ keV$\cdot $b. The important role of the asymptotic
			constant (AC) for the $^{15}$N($p,\gamma _{0}$)$^{16}$O process with
			interfering $^{3}S_{1}$(312) and $^{3}S_{1}$(962) resonances is elucidated.
			A comparison of our calculation for $S-$factor with existing experimental
			and theoretical data is addressed, and a reasonable agreement is found.
			
			The reaction rate is calculated and compared with the existing rates. It has
			negligible dependence on the variation of AC, but shows a strong impact of
			the interference of $^{3}S_{1}$(312) and $^{3}S_{1}$(962) resonances,
			especially at temperatures, referring to the CNO Gamow windows. We estimate
			the contribution of cascade transitions to the reaction rate based on the
			exclusive experimental data by \textit{Imbriani, et al. 2012}. The reaction
			rate enhancement due to the cascade transitions is observed from $T_{9} >
			0.3 $ and reaches the maximum factor $\sim $\ 1.3 at $T_{9}=1.3$.
			
			We present the Gamow energy window and a comparison of rates for radiative
			proton capture reactions $^{12}$N($p,\gamma $)$^{13}$O, $^{13}$N($p,\gamma $)%
			$^{14}$O, $^{14}$N($p,\gamma $)$^{15}$O, and $^{15}$N($p,\gamma $)$^{16}$O
			obtained in the framework of the MPCM and give temperature windows,
			prevalence, and significance of each process.
		\end{abstract}
		
		
	\maketitle

\section{Introduction}

Stellar burning depend on the star's initial mass and can proceed either
through the $p-p$ chain or through the Carbon-Nitrogen-Oxygen (CNO) cycle,
fusing hydrogen to helium through a chain fusion processes, $-$ sequence of
thermonuclear reactions that provides most of the energy radiated by the hot
stars \cite{Barnes1982, Adelberger2011,Arnould2020}. Unlike the $p-p$ chain,
the CNO cycle is a catalytic one, that converts 4 protons into one helium
nucleus but does so via reactions on the pre-existent seed isotopes of
carbon, nitrogen, and oxygen nuclei. The carbon, nitrogen, and oxygen
isotopes act just as catalysts in the CNO cycle.
The CNO\ bi-cycle involves the following chains of nuclear reactions:

\begin{equation}
	^{\text{12}}\text{C(}p,\gamma \text{)}^{\text{13}}\text{N(}\beta ^{+}\nu
	\text{)}^{\text{13}}\text{C(}p,\gamma \text{)}^{\text{14}}\text{N(}p,\gamma
	\text{ )}^{\text{15}}\text{O(}\beta ^{+}\nu \text{)}^{\text{15}}\text{N}%
	\begin{tabular}{c}
		\ \ \ \ \ \ $\longrightarrow \text{ }^{\text{15}}\text{N(}p,\alpha \text{)}^{%
			\text{12}}\text{C \ \ \ \ \ \ \ \ \ \ \ \ \ \ \ \ \ \ \ \ \ \ \ \ \ \ \ \ \
			\ \ \ \ \ \ \ \ \ \ \ \ \ }$ \\
		$\ \ \ \ \ \ \ \longrightarrow \text{ \ }^{\text{15}}\text{N(}p,\gamma \text{)}^{\text{16}}%
		\text{O(}p,\gamma \text{)}^{\text{17}}\text{F(}\beta ^{+}\nu \text{)}^{\text{%
				17}}\text{O(}p,\alpha \text{)}^{\text{14}}\text{N}$.%
	\end{tabular}
	\label{Limit}
\end{equation}%

Therefore, the CNO bi-cycle produces three electron neutrinos from $\beta
^{+}$ decay of $^{\text{13}}$N, $^{\text{15}}$O, and $^{\text{17}}$F and is
also referred to as the \textquotedblleft cold\textquotedblright\ CNO cycle
\cite{Wiescher2010}. The CN cycle contains no stable $^{\text{13}}$N and $^{%
	\text{15}}$O isotopes, that decay to the stable isotopes $^{\text{13}}$C and
$^{\text{15}}$N, respectively. \ The catalytic nuclei are lost from the
process via the leak reaction $^{\text{15}}$N($p,\gamma $)$^{\text{16}}$O
and the subsequent reactions in (\ref{Limit}) restore the catalytic
material, generating $^{\text{16}}$O and heavier isotopes leading to the
accumulation of the\ $^{\text{4}}$He and $^{\text{14}}$N nuclei. This second
branch produces $^{\text{17}}$F, which $\beta ^{+}$ decay with the emission
of the 1.74 MeV electron neutrinos. Thus, the $^{\text{15}}$N($p,\gamma $)$^{%
	\text{16}}$O process represents a branching reaction linking the alternative
NO channel of the CNO cycle that produces the stable oxygen isotopes \cite%
{deBoer2013}. Therefore, in the CNO cycle, the proton capture reaction on $%
^{15}$N allows two possible channels: the branch of the cycle\ $^{15}$N($%
p,^{4}$He)$^{12}$C and the branch of the cycle $^{15}$N($p,\gamma $)$^{16}$%
O, reactions and they intersect at the $^{\text{15}}$N nucleus.

The rate of the CN with respect to the NO cycle depends on the branching
ratio of the $^{\text{15}}$N($p,\gamma $)$^{\text{16}}$O and $^{\text{15}}$N(%
$p,\alpha $)$^{\text{12}}$C reaction cross sections. The probability for the
$^{\text{15}}$N($p,\gamma $)$^{\text{16}}$O process to occur is about one
for every thousand of the second \cite{Caciolli2011}, thus the contribution
to the overall nuclear energy production is negligible, while the
consequences on the nucleosynthesis are critical \cite{Boeltzig2016}.
Therefore, in the case of an active NO cycle, the correct evaluation of the $%
^{\text{15}}$N($p,\gamma $)$^{\text{16}}$O reaction is crucial to properly
predict the abundances of all the stable $^{\text{16}}$O, $^{\text{17}}$O,
and $^{\text{18}}$O isotopes and their relative ratios \cite%
{Caughlan1962,Caughlan1983,Caughlan1988}. The reaction rates ratio
determines on how much nucleosynthesis of $^{\text{16}}$O, $^{\text{17}}$O,
and $^{\text{18}}$O takes place during CNO burning \cite{Caughlan1962}.

Since the first experimental study of $^{15}$N($p,\gamma $)$^{16}$O reaction
in 1952 \cite{Schardt1952} experimental data \cite%
{Hebbard,Brochard1973,Rolf1974,Bemmerer2009,LeBlanc2010,Imbriani2012,Caciolli2011}
for total cross sections of the radiative $p^{15}$N capture in the energy
region from 80 keV to 2.5 MeV were collected \cite{Angulo1999,Xu2013}.
Analysis of existing experimental measurements of the low-energy $^{\text{15}%
}$N($p,\gamma $)$^{\text{16}}$O reaction shows that cross section data
differ substantially at lower energies.

In the past, a variety of theoretical approaches from potential cluster
models to multilevel $R-$matrix formalisms \cite%
{Barker2008,Mukhamedzhanov2008,LeBlanc2010,Mukhamedzhanov2011,deBoer2013,Dubovichenko2014}
were used to describe the $^{\text{15}}$N($p,\gamma $)$^{\text{16}}$O
reaction cross section at the stellar energies and astrophysical $S-$factor
that is the main characteristic of this process at low energies. In the
framework of the selective resonant tunneling model \cite{Tian2000} $^{\text{%
		15}}$N($p,\gamma $)$^{\text{16}}$O cross section and $S-$factor have been
studied \cite{Khan2022}. Most recently, the astrophysical $S-$factor for the
radiative proton capture process on the $^{15}$N nucleus at stellar energies
are studied within the framework of the cluster effective field theory \cite%
{SonNP2022,Son2022}. The authors perform the single channel calculations
where only the contribution of the first resonance into a cross section was
considered \cite{SonNP2022} and then reported the results by including two
low-energy resonances \cite{Son2022}.

In this paper we are continuing the study of the reactions of radiative
capture of protons on light atomic nuclei \cite{PRCDTKB2020,PRC2022DTKB} and
consider the radiative proton capture on $^{15}$N at astrophysical energies
in the framework of a modified potential cluster model (MPCM). Within the
MPCM over thirty reactions of radiative capture of protons, neutrons and
other charged particles on light atomic nuclei were considered and results
are summarized in \cite{DubBook2015,DubBook2019}. References \cite%
{PRCDTKB2020,PRC2022DTKB} provide the basic theoretical framework of the
MPCM approach for the description of a charged-particle-induced radiative
capture reactions. Calculation methods based on the MPCM of light nuclei
with forbidden states (FS) are used \cite{Neudatchin1992}. The presence of
allowed states (AS) and FS are determined based on the classification of the
orbital states of clusters according to Young diagrams \cite{Kukulin1983}. 
Our analysis is data driven: the potentials of intercluster
interactions for scattering processes are constructed based on the
reproduction of elastic scattering phase shifts, taking into account their
resonance behavior or the spectra of the levels of the final nucleus. For
the bound states (BS) or ground state (GS) of nuclei in cluster channels,
intercluster potentials are built based on a description of the binding
energy, asymptotic normalization coefficient (ANC) and mean square radius
\cite{DubBook2019,PRCDTKB2020,PRC2022DTKB}.

The modified model includes a classification of orbital states according to Young's 
diagrams. This classification leads to the concept of “forbidden states” in interaction 
potentials. In particular, for the GS of the $p^{15}$N system, the potential has a deeply 
bound state, which is the FS, and the second bound state is the GS of the $^{16}$O 
nucleus in this channel. The concept of forbidden states makes it possible to effectively 
take into account the Pauli principle for multinucleon systems when solving problems 
in cluster models for a single-channel approximation. The current study is based on 
the detailed classification of orbital states in $p \ + ^{15}$N channel by Young's diagrams 
we performed early in Ref.  \cite{Dubovichenko2014}.

In this work we study a radiative $p^{15}$N capture on the ground state of $%
^{16}$O within the framework the MPCM. For the first time the contribution
of $^{3}P_{1}$ scattering wave in $p$ + $^{15}$N channel due to $%
^{3}P_{1}\longrightarrow $ $^{3}P_{0}$ $M1$ transition is considered. Our
approach allows to analyze the explicit contribution of each transition into
the $S-$factor and show the origin of the interference for $E1$ transitions.

This paper is organized as follows. Section II presents a structure of
resonance states and construction of interaction potentials based on the
scattering phase shifts, mean square radius, asymptotic constant and bound
states or ground state of $^{16}$O nucleus. Results of calculations of the
astrophysical $S-$factor and reaction rate for the proton radiative capture
on $^{15}$N are presented in Secs. III and IV, respectively. In the same
sections we discuss a comparison of our calculation for $S-$factor and
reaction rate with existing experimental and theoretical data. Conclusions
follow in Sec. V.

\section{Interaction potentials and structure of resonance states}

The $E1$ transitions from resonant $^{3}S_{1}-$scattering states are the
main contributions to the total cross section of the radiative proton
capture on $^{15}$N to the ground state of $^{16}$O \cite{deBoer2013}. In
the channel of $p$ + $^{15}$N in continuum there are two $^{3}S_{1}$
resonances:

1.\qquad The first resonance appears at an energy of 335(4) keV with a width
of 110(4) keV in the laboratory frame and has a quantum numbers $J^{\pi }$, $%
T=1^{-}$, $0$ (see Table 16.22 in \cite{Ajzenberg1993}). This resonance is
due to the triplet $^{3}S_{1}$ scattering state and leads to $E1$ transition
to the GS. In the center-of-mass (c.m.) frame this resonance is at an energy
of 312(2) keV with a width of 91(6) keV and corresponds to the resonant
state of the $^{16}$O at an excitation energy of $E_{\text{x}}=12.440(2)$
MeV (see Table 16.13 in \cite{Ajzenberg1993}). However, in the new database
\cite{Sukhoruchkin2016}, for this resonance, the excitation energy of $E_{%
	\text{x}}=12.445(2)$ MeV and the width of $\Gamma =101(10)$ keV in the c.m.
are reported.

2.\qquad The second resonance is at an energy of 1028(10) keV with a width
of 140(10) keV in laboratory frame and has a quantum numbers $J^{\pi }=1^{-}$
and $T=1$ \cite{Ajzenberg1993}. This resonance is also due to the triplet $%
^{3}S_{1}$ scattering and leads to $E1$ transition to the GS of $^{16}$O. In
the c.m. the resonance emerges at an energy of 962(2) keV with a width of $%
\Gamma =139(2)$ keV and corresponds to the excitation energy of $E_{\text{x}%
}=13.090(2)$ MeV of $^{16}$O in a new database \cite{Sukhoruchkin2016}. In
the database \cite{Ajzenberg1993} for this resonance, the excitation energy
of $E_{\text{x}}=13.090(8)$ MeV and width of $\Gamma =130(5)$ keV in the
c.m. are reported.

\begin{table}[h]
	\caption{Data on the $^{3}S_{1}$ resonance states in $p$ + $^{15}$N channel.
		$E_{\text{x}}$ is the excitation energy, $E_{\text{res}}$ and $\Gamma _{%
			\text{res}}$ are the experimental resonance energy and the width,
		respectively. $E_{\text{theory}}$ and $\Gamma _{\text{theory}}$ are the
		resonance energy and the width, respectively, obtained in the present
		calculations.}
	\label{tab:Table1}
	\begin{center}
		\begin{tabular}{cccccc}
			\hline
			$^{2S+1}L_{J}$ & $E_{\text{x}}$, MeV & $E_{\text{res}}$, keV & $\Gamma _{%
				\text{res}}$, keV & $E_{\text{theory}}$, keV & $\Gamma _{\text{theory}}$, keV
			\\ \hline
			$^{3}S_{1}$(312) &
			\begin{tabular}{c}
				12.440(2) \cite{Ajzenberg1993} \\
				12.445(2)\cite{Sukhoruchkin2016}%
			\end{tabular}
			&
			\begin{tabular}{c}
				312(2) \cite{Ajzenberg1993} \\
				317(2) \cite{Sukhoruchkin2016}%
			\end{tabular}
			&
			\begin{tabular}{c}
				91(6) \cite{Ajzenberg1993} \\
				101.5(10) \cite{Sukhoruchkin2016}%
			\end{tabular}
			& 312 & $125-141$ \\
			$^{3}S_{1}$(962) &
			\begin{tabular}{c}
				13.090(8) \cite{Ajzenberg1993} \\
				13.090(2) \cite{Sukhoruchkin2016}%
			\end{tabular}
			&
			\begin{tabular}{c}
				962(8) \cite{Ajzenberg1993} \\
				962(2) \cite{Sukhoruchkin2016}%
			\end{tabular}
			&
			\begin{tabular}{c}
				130(5) \cite{Ajzenberg1993} \\
				139(2) \cite{Sukhoruchkin2016}%
			\end{tabular}
			& 962 & $131$ \\ \hline
		\end{tabular}%
	\end{center}	
\end{table}

The compilation of experimental data on the $^{3}S_{1}$ resonances is
presented in Table \ref{tab:Table1}.

In databases \cite{Ajzenberg1993,Sukhoruchkin2016} are reported the other
resonances as well. The third resonance has an energy of 1640(3) keV with a
width of 68(3) keV in the laboratory frame and quantum numbers $J^{\pi }$,$%
T=1^{+},0$. This resonance can be due to the triplet $^{3}P_{1}$ scattering
and leads to $M1$ transition to the GS. The resonance is at an energy of
1536(3) keV with a width of 64(3) keV in the c.m. that corresponds to the
excitation energy 13.664(3) MeV of the $^{16}$O \cite{Ajzenberg1993} and in
Ref. \cite{Sukhoruchkin2016} the excitation energy of 13.665(3) MeV and the
width of 72(6) keV in the c.m. are reported. However, this resonance was
observed only in measurements \cite{Rolf1974}, and in the later measurements
\cite{LeBlanc2010,Imbriani2012} the resonance is absent. Therefore, we will
not consider it in our calculations. The next resonance is excited at the
energy of 16.20(90) MeV ($J^{\pi }$,$T=1^{-},0$) has a larger width of
580(60) keV in the c.m. and its contribution to the reaction rate will be
small. In addition, in the spectra of $^{16}$O \cite{Ajzenberg1993}, another
resonance is observed at an excitation energy of 16.209(2) MeV ($J^{\pi }$,$%
T=1^{+},1$) with a width of 19(3) keV in the c.m. However, the resonance
energy is too large and its width too small to make a noticeable
contribution to the reaction rates.

The cascade transitions via two very narrow 2$^{-}$ resonances as well as
the 0$^{-}$ and 3$^{-}$ resonances in the 0.40 $\lesssim E_{R}\lesssim $
1.14 MeV range \cite{Gorodetszky1968,Imbriani2012,Imbriani2012Err} are
considered for the reaction rate calculations. These cascading transitions
are included in the NACRE II \cite{Xu2013} reaction rate calculations and
appear at high $T_{9}$. There are two 2$^{-}$ resonances \cite{Imbriani2012}%
, at the excitation energies of 12.53 Mev and 12.9686 MeV with the widths of
97(10) eV and 1.34(4) keV \cite{Sukhoruchkin2016}, respectively. When one
considers transitions to the ground state with 0$^{+}$, only $M2$
transitions are possible here, which have a very small cross section, and we
do not consider them. In addition, due to such small widths, their
contribution to the reaction rate will be very small. The measurement of the
excitation functions of the three dominant cascade transitions allows one to
estimate of the contributions from these transitions. In Ref. \cite%
{deBoer2013} capture processes to the GS and three excited states $E_{\text{x%
}}$ $=6.049$ MeV, 6.130 and 7.117 MeV are considered, that gives $S(0)=$
41(3) {keV}${\cdot }${b} in total while 40(3) {keV$\cdot $b} for the GS. It
is shown that the 1$^{-}$ and two 2$^{+}$ resonances do not affect the value
of the $S-$factor, two 3$^{-}$ resonances at $E_{\text{x}}$ = 13.142 MeV and
13.265 MeV decay into the GS due to the $E3$ transition and their
contribution is negligible \cite{deBoer2013}.

On the background of strong $E1$ transitions the next one in the
long-wave expansion of the electromagnetic Hamiltonian is magnetic dipole $%
M1 $ transition \cite{Greiner}. In the case of $^{15}$N($p,\gamma _{0}$)$%
^{16}$O, $M1$ transition is allowed by selection rules. It occurs as a
direct capture from the non-resonance scattering wave $^{3}P_{1}$ to the $%
^{16}$O $^{3}P_{0} $ state. The intensity of the $M1$ partial transition
depends on the distorted $^{3}P_{1}$ wave only and is related to the
corresponding interaction potential. Our estimations of the $M1$ partial
cross section in plane-wave approximation shows near order of magnitude
suppression. The inclusion of the $p-$wave interaction in $p+^{15}$N channel
enhances the $M1$ transition. The interaction potential could be constructed
based on an elastic $^{15}$N($p,p$)$^{15}$N scattering to describe the $%
^{3}P_{1}$ phase shift. The phase shifts should satisfy the following
conditions: i. at $E=0$ the $\delta _{^{3}P_{1}}=180^{0}$ according to the
generalized Levinson theorem \cite{Neudatchin1992}; ii. fit the exsisting $p-
$wave $^{15}$N($p,p$)$^{15}$N scattereng data. iii. have a non-resonace
behaviour. Below we found potential parameters that provides a reasonable
phase shifts. Therefore, in calculations we are considering only the above
two $^{3}S_{1}$ resonance transitions and non-resonance $^{3}P_{1}$
scattering for the $M1$\ transition to the $^{16}$O GS.

We should distinguish $M1$ capture from the non-resonance $%
^{3}P_{1}$ scattering wave to the $^{16}$O $^{3}P_{0}$ state and $M1$
de-excitation of 1$^{+}$ level at 13.665 MeV to the ground state. As pointed
by \textit{deBoer et al.} {\cite{deBoer2013}}: \textquotedblleft The 1$^{+}$
level at $E_{x}$ = 13.66 MeV could decay by $M1$ de-excitation to the ground
state but no evidence for this is observed (contrary to Ref. {\cite{Rolf1974}%
}).\textquotedblright\ As mentioned above we are not considering the
contribution of $M1$ de-excitation (1$^{+}$ level at 13.665 MeV) to the $%
S(E) $.

For calculations of the total radiative capture cross sections, the nuclear
part of the $p^{15}$N interaction potential is contracted using Gaussian
form \cite{DubBook2019,PRCDTKB2020,PRC2022DTKB}:
\begin{equation}
	V(r,JLS,\{f\})=-V_{0}(JLS,\{f\})exp\left( -\alpha (JLS,\{f\})r^{2}\right) .
	\label{potential}
\end{equation}%
The parameters $\alpha $ and $V_{0}$ in Eq. (\ref{potential}) are the
interaction range and the strength of the potential, respectively.

\begin{table}[h]
	\caption{Parameters of interaction potentials $V_{0}$ and $\protect\alpha$
		for the GS and continuum states. The $C_{W}$ is a dimensionless constant
		that corresponds to the range of the experimental ANC $12.88 - 14.76$ fm$%
		^{-1/2}$ \protect\cite{Mukhamedzhanov2008,Mukhamedzhanov2011}. The
		theoretical widths, $\Gamma _{\text{theory}}$, for the resonance $^{3}S_{1}$%
		(312) and $^{3}S_{1}$(962) are calculated using the corresponding parameters
		of the potentials. $V_{0}$ and $\Gamma _{\text{theory}}$ are given in MeV
		and keV, respectively, $\protect\alpha$ in fm$^{-2}$, and ANC in fm$^{-1/2}$%
		. }
	\label{tab:Table1_1}
	\begin{center}
		\begin{tabular}{c|cccc|ccc|ccc|cc}
			\hline
			& \multicolumn{4}{|c}{$^{3}P_{0}$, GS} & \multicolumn{3}{|c|}{$^{3}S_{1}$%
				(312), $E1$} & \multicolumn{3}{|c|}{$^{3}S_{1}$(962), $E1$} &
			\multicolumn{2}{|c}{$^{3}P_{1}$, $M1$} \\ \cline{2-13}
			Set & $V_{0}$ & $\alpha $ & ANC & $C_{W}$ & $V_{0}$ & $\alpha $ & $\Gamma _{%
				\text{theory}}$ & $V_{0}$ & $\alpha $ & $\Gamma _{\text{theory}}$ & $V_{0}$
			& $\alpha $ \\ \hline
			I & 976.85193 & 1.1 & 14.49 & 2.05 & 1.0193 & 0.0028 & 141 &  &  &  &  &  \\
			II & 1057.9947 & 1.2 & 13.71 & 1.94 & 1.0552 & 0.0029 & 131 & 105.0675 & 1.0
			& 131 & 14.4 & 0.025 \\
			III & 1179.3299 & 1.35 & 12.85 & 1.8 & 1.0902 & 0.003 & 125 &  &  &  &  &
			\\ \hline
		\end{tabular}%
	\end{center}
\end{table}

The strength and the interaction range of the potential (\ref{potential})
depend on the total and angular momenta, the spin, $JLS$, and Young diagrams
$\{f\}$ \cite{DubBook2015,DubBook2019}. For description of the $^{3}S_{1}$
scattering states we use the corresponding experimental energies and widths {%
	from Table \ref{tab:Table1}. } Coulomb potential is chosen as point-like interaction potential.

Construction of the potentials that give the energies and widths of $%
^{3}S_{1}$(312) and $^{3}S_{1}$(962) resonances reported in the literature
is a challenging task. 
One has to find the optimal parameters of the potentials for the description
of $E1$ transitions that lead to the fitting of the experimental resonance
energies and the widths of both interfering resonances. For the $^{3}S_{1}$%
(962) resonance the found optimal parameters of the interaction potential
that allow {to reproduce the resonance energy $E_{\text{res}}=962(1)$ keV
	and width $\Gamma _{\text{res}}=131$ keV are }reported in {Table \ref%
	{tab:Table1_1}.} The situation is more complicated with the $^{3}S_{1}$(312)
resonance. %
While it is possible to reproduce rather accurately the position and the
width of the $^{3}S_{1}$(312) resonance, the consideration of the
interference of $^{3}S_{1}$ resonances gives different sets of optimal
parameters for the potential. We found three sets I $-$ III of the optimal
values for $V_{0}$ and $\alpha $ parameters reproducing exactly the energy
of the first resonance $E_{\text{theory}}=312(1)$ keV, but the width $\Gamma
_{\text{theory}}$ varies in the range of $125-141$ keV.

\begin{figure}[h]
	\centering
	\includegraphics[width=8.0cm]{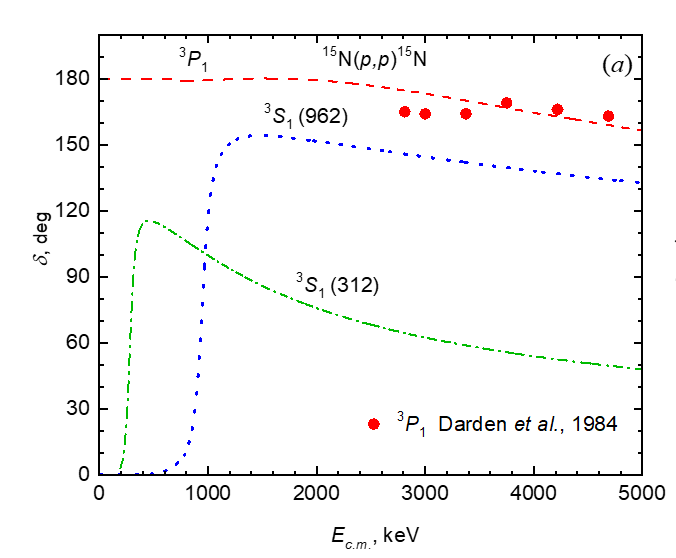} \includegraphics[width=7.5cm]{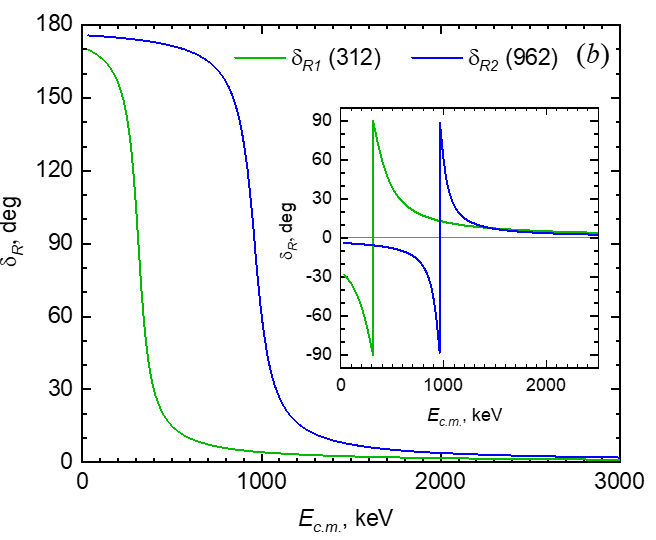}
	\caption{(Color online) The dependence of the elastic $p^{15}$N scattering
		phase shifts on the energy. ($a$) Calculations are performed by using the
		potentials with parameters from Table \protect\ref{tab:Table1_1}. The phase
		shifts for $^{3}S_{1}$(312) resonance is calculated using the set I and is
		shown by the dash-dotted curve. The phase shifts for $^{3}S_{1}$(962)
		resonance and $^{3}P_{1}$ are presented by the dotted and dashed curves,
		respectively. The experimental data from Ref. \protect\cite{DARDEN1984}. ($b$%
		) The energy dependence of the resonant phase shifts $\protect\delta _{R1}$
		and $\protect\delta _{R2}$ calculated using the experimental resonance
		widths 91 keV and 130 keV, respectively. Reconstructed scattering resonant
		phase shifts $\protect\delta _{R1}$ and $\protect\delta _{R2}$ are
		calculated using Eq. (\protect\ref{DeltaR}) and taking into account the
		tangent function periodicity. The insert shows the result of calculation
		using Eq. (\ref{DeltaR}). }
	\label{Fig1}
\end{figure}

The dependence of the elastic $p^{15}$N scattering phase shifts for the $E1$
transitions on the energy is shown in Fig. \ref{Fig1}$a$. The result of the
calculation of the $^{3}S_{1}$ phase shifts with the parameters for the $S$
scattering potential without FS from Table \ref{tab:Table1_1} leads to the
value of 90$^{\circ }$(1) at the energies 312(1) and 962(1) keV,
respectively. Our calculations of the phase shifts for the $^{3}S_{1}$(312)
resonance with the parameter sets I - III show very close energy dependence
in the entire energy range up to 5 MeV at a fixed resonance position.

In elastic $p^{15}$N scattering spectra at energies up to 5 MeV, there are
no resonance levels with $J^{\pi }=0^{+},1^{+},2$ except for the mentioned
above, and widths greater than 10 keV \cite{Ajzenberg1993}. Therefore, for
potentials of non-resonance $^{3}P-$waves with one bound FS parameters can
be determined based on the assumption that in the energy region under
consideration their phase shifts are practically zero or have a gradually
declining character \cite{PRC2022DTKB}. For such potential the optimal
parameters are: $V_{p}=14.4$ MeV$,$ $\alpha _{p}=0.025$ fm$^{\text{-2}}$.
The result of the calculation of $P-$phase shifts with such potential at an
energy of up to 5 MeV is shown in Fig. \ref{Fig1}$a$. To determine the
values of phase shifts at zero energy, we use the generalized Levinson
theorem \cite{Neudatchin1992}, so the phase shifts of the potential with one
bound FS should begin from 180$^{\circ }$. In the energy region $E_{\text{%
		c.m.}}<5$ MeV the $^{3}P_{1}$ phase shift has gradual energy dependence and
it is almost constant up to $E_{\text{c.m.}}\lesssim 2.2$ MeV. \

It is interesting to compare experimentally determined phase shifts with our
calculations. While in Ref. \cite{DeBoer2012} the elastic scattering of
protons from $^{15}$N was studied and authors measured the excitation
functions of $^{15}$N($p,p$)$^{15}$N over the proton energy range from 0.6
to 1.8 MeV at some laboratory angles, there been no phase shifts reported.
There has been no systematic experimentally determined phase shifts at the
energies of the astrophysical interest. Absolute differential cross sections
were measured for the reactions $^{15}$N($p,p$)$^{15}$N \cite{Bashkin1959}
and spins and parities are discussed where resonances suggest the existence
of excited states in $^{16}$O. In Ref. \cite{LaCanna1976} authors carried
out a phase-shift analysis of cross section for the energy interval 8$-$15
MeV\textbf{.} Angular distributions of cross section and analyzing power for
elastic scattering of protons from $^{15}$N have been measured for the
energy interval 2.7$-$7 MeV \cite{DARDEN1984}, and the authors gave a
phase-shift analysis of data. Results of our calculations for
the $^{3}P_{1}$ phase shift along with the experimental data \cite%
{DARDEN1984} are presented in Fig. \ref{Fig1}$a$.

The main assumption in all previous studies of the reaction $^{15}$N($%
p,\gamma $)$^{16}$O was that a direct and resonant radiative capture cross
sections and their interferences contribute to the total cross section. The
direct radiative capture process is considered assuming a potential
peripheral radiative capture for a hard sphere scattering \cite%
{Rolfs1973,Rolf1974,Rolf19742,Blokhintsev2022,Mukhamedzhanov2023}. However,
in general, phase shifts are extracted from experimental data analysis
without the separation on the potential and resonant terms \cite%
{Davydov,Iliadis2015}. We follow this Ansatz. Consequently, the $E1$
transition amplitudes are constructed based on a single radial scattering
wave function resembling continuous, both smooth and resonance energy
dependence. Both $^{3}S_{1}$ phase shifts obtained with these scattering
wave functions depend on the energy as shown in Fig. \ref{Fig1}$a$.
Therefore, we have only the interference between two partial $E1$ matrix
elements in contrast to all previous considerations.

The resonant phase shift is given by the usual expression \cite%
{Rolf1974,Iliadis2015}
\begin{equation}
	\delta _{R}=\tan ^{-1}\frac{\Gamma }{2(E-E_{res})},  \label{DeltaR}
\end{equation}%
where $\Gamma $ is the width of a resonance. In Fig. \ref{Fig1}$b$ the
energy dependence of the resonant $\delta _{R1}$\ and $\delta _{R2}$ phase
shifts for the resonance width 312 keV and 962 keV is presented,
respectively. The comparison of the phase shifts for the $E1$ transitions
via the $^{3}S_{1}$(312) and $^{3}S_{1}$(962) resonances with the resonant
phase shifts $\delta _{R1}$\ and $\delta _{R2}$ shows their different energy
dependence.

We construct the potential for $^{16}$O in GS with \textit{J}$^{\pi }$%
\textit{,T} = 0$^{+}$,0 in $p^{15}$N-channel based on the following
characteristics: the binding energy of 12.1276 MeV, the experimental values
of 2.710(15) fm and 2.612(9) fm \cite{Ajzenberg1993} for the root mean
square radii of $^{16}$O and $^{15}$N of \cite{Ajzenberg1991}, respectively,
and a charge and matter radius of a proton 0.8414 fm \cite{Website1}. The
potential also should reproduce the AC. The corresponding potential includes
the FS and refers to the $^{3}P_{0}$ state.

Usually for a proton radiative capture reaction of astrophysical interest
one assumes that it is peripheral, occurring at the surface of the nucleus.
If the nuclear process is purely peripheral, then the final bound-state wave
function can be replaced by its asymptotic form, so the capture occurs
through the tail of the nuclear overlap function in the corresponding
two-body channel. The shape of this tail is determined by the Coulomb
interaction and is proportional to the asymptotic normalization coefficient.
The role of the ANC in nuclear astrophysics was first discussed by
Mukhamedzhanov and Timofeyuk \cite{Timofeyuk1990} and in Ref. \cite{Xu1994}.
These works paved the way for using the ANC approach as an indirect
technique in nuclear astrophysics. See Refs. \cite%
{Timofeyuk1995,Mukhamedzhanov2001,Timofeyuk2003,Mukhamedzhanov2003,Timofeyuk2009,Timofeyuk2013,Mukhamedzhanov2014,Blokhintsev2021,Blokhintsev2022}
and citation herein and the most recent review \cite{Mukhamedzhanov2023}.

We construct a potential with the FS $^{3}P_{0}$ state using the
experimental ANC given in Ref. \cite{Mukhamedzhanov2008} that relates to the
asymptotics of radial wave function as $\chi _{L}(R)=CW_{-\eta
	L+1/2}(2k_{0}R)$. The dimensional constant $C$ is linked with the ANC via
the spectroscopic factor $S_{F}$. In our calculations we exploited the
dimensionless constant \cite{Plattner1981} $C_{W}$, which is defined in \cite%
{PRC2022DTKB} {as $C_{W}=C/\sqrt{2k_{0}}$, where $k_{0}$ is wave number
	related to the binding energy}. In Ref. \cite{Mukhamedzhanov2008} the values$%
\ 192(26)$ fm$^{-1}$ and $2.1$ were reported for the ANC\ and spectroscopic
factor, respectively. In Ref. \cite{Mukhamedzhanov2011} the dimensional ANC
includes the antisymmetrization factor $N$ into the radial overlap function
as it was clarified by authors in \cite{Mukhamedzhanov2014}. The factor $N$
is defined as $N=\left(
\begin{matrix}
	A \\
	x%
\end{matrix}%
\right) ^{1/2}=\sqrt{\dfrac{A!}{(A-x)!x!}},$ where ${x}$ and ${A}$ are the
atomic mass numbers of the constituent nucleus from ${x}$ and ${A-x}$ nucleons,
respectively \cite{Blokhintsev1977}. If $x=1$, then $N=\sqrt{A}$ and for the
reaction $^{15}$N($p,\gamma _{0}$)$^{16}$O $N=4$. Thus, using
the experimental square of the ANC $192\pm 26$ fm$^{-1}$ \cite%
{Mukhamedzhanov2008,Mukhamedzhanov2011}, we obtained the interval for the
dimensionless AC used in our calculations: $C_{W}=1.82-2.09$ that
corresponds to the ANC of $12.88-14.76$ fm$^{-1/2}$. In the present
calculations, we use for the proton mass $m_{p}=1.00727646677$ amu \cite%
{Website1}, $^{15}$N mass 15.000108 amu \cite{Website2}, and the constant $%
\hbar ^{2}/m_{0}=41.4686$ MeV\textperiodcentered fm$^{2}$, where $%
m_{0}=931.494$ MeV is the atomic mass unit (amu).

The $^{15}$N($p,\gamma $)$^{16}$O is the astrophysical radiative capture
process, in which the role of the ANC is elucidated \cite{Mukhamedzhanov2023}%
. In Table \ref{tab:Table1_1} three sets of parameters for the $^{3}P_{0}$
GS potential and AC $C_{W}$ are listed. The asymptotic constant $C_{W}$ is
calculated over averaging at the interval $5-10$ fm. Each set leads to the
binding energy of 12.12760 MeV, the root mean square charge radius of 2.54
fm and the matter radius of 2.58 fm, but the sets of $C_{W}$ lead to the
different widths of the $^{3}S_{1}$(312) resonance.

Note, that there is one important benchmark for the choice of optimal sets
for the parameters of interaction potentials for the first $E1$(312)
resonance. There are the experimental values of the total cross section $%
\sigma _{\text{exp}}(312)=6.0\pm 0.6$ $\mu $b \cite{Caciolli2011} and $%
6.5\pm 0.6$ $\mu $b \cite{LeBlanc2010}, which are in excellent agreement
with earlier data $6.3$ $\mu $b \cite{Brochard1973} and $6.5\pm 0.7$ $\mu $b
\cite{Hebbard}. Simultaneous variation of $C_{W}$ for the GS and parameters $%
V_{0}$ and $\alpha $ for the $^{3}S_{1}$(312) was implemented to keep the
value of the cross section $\sigma _{\text{theory}}(312)=5.8-5.9$ $\mu $b
matching the experimental data. The result of this optimization is presented
in Table \ref{tab:Table1_1} as sets I $-$ III.

Table \ref{tab:Table1_1} summarizes the potential parameters used in the
case where the MPCM works reasonably well for a radiative proton capture in
the $^{\text{15}}$N($p,\gamma _{0}$)$^{\text{16}}$O reaction.

We sum up the procedure and choice of potential parameters as:

1.\quad we construct the nucleon-nuclei potentials that give the channel
binding energy with the requested accuracy 10$^{-5}$ MeV. There are a few
such potentials;

2.\quad the experimental ANC is used as a criterion for the choice of the
potential that provides the required asymptotic behavior of the radial wave
function at the fixed binding energy. Thus, the variety of wave function is
constrained within the upper and lower limits for the ANC: $12.88-14.76$ fm$%
^{-1/2}$ for $^{15}$N($p,\gamma $)$^{16}$O;

3.\quad an additional test of the wave functions is a reproduction of the
matter and charge radii with a precision of $\sim $ 5\% and the $\sigma _{%
	\text{exp}}(312)$ cross section within experimental uncertainties;

4.\quad for the continuous spectrum the parameters of the potential are
fixed by the resonance energy and width above threshold. An additional
source of the $S-$factor uncertainty relates to uncertainties of the
resonance energy and width;

5.\quad this procedure gives the model's uncertainty bands for the $S-$%
factor. 

\section{Astrophysical $S-$Factor}

The astrophysical $S-$factor is the main characteristic of any thermonuclear
reaction at low energies. The present analysis focuses primarily on
extrapolating the low-energy $S-$factor of the reaction $^{15}$N($p,\gamma $)%
$^{16}$O into the stellar energy range. Since the first experimental study
of $^{15}$N($p,\gamma _{0}$)$^{16}$O reaction in 1960 \cite{Hebbard},
experimental data \cite%
{Rolf1974,LeBlanc2010,Imbriani2012,Caciolli2011,Xu2013} for total cross
sections of the radiative $p^{15}$N capture in the energy region from 80 keV
to 2.5 MeV have been collected. These experimental studies verified and
confirmed that the radiative $p^{15}$N capture is dominated by the first two
interfering resonances at 312 keV and 962 keV with the quantum numbers $%
J^{\pi }$, $T=1^{-}$,$0$ and $J^{\pi }$, $T=1^{-}$,$1$, respectively.

\subsection{$E1$ transitions}

The $E1$ transitions are the main input parts of the radiative capture
amplitude for $^{15}$N($p,\gamma _{0}$)$^{16}$O reaction. Therefore, it is
required to determine the resonance capture cross sections for these
transitions accurately to avoid one of the main sources of uncertainty. The
radiative resonance capture to the bound states is reviewed in Ref. \cite%
{Mukhamedzhanov2023}. Following Ref. \cite{PRC2022DTKB} after algebraic
calculations using quantum numbers related to the $^{15}$N($p,\gamma $)$%
^{16} $O reaction, one can write the cross section for the radiative capture
$p^{15}$N to the ground state of $^{16}$O as

\begin{equation}
	\sigma _{E1}(E_{\text{c.m.}})=\frac{4\pi e^{2}}{9\hbar ^{2}}\left( \frac{K}{k%
	}\right) ^{3}\left( \frac{1}{m_{p}}-\frac{7}{m_{^{15}\text{N}}}\right)
	^{2}\left\vert I(k;E1)\right\vert ^{2}.  \label{CrossSec}
\end{equation}%
In Eq. (\ref{CrossSec}) $\mu $ is the reduced mass of the proton and $^{15}$%
N nucleus, $K=E_{\gamma }/\hbar c$ is the wave number of the emitted photon
with energy $E_{\gamma }$, $k$ is relative motion wave number and

\begin{equation}
	\left\vert I(k;E1)\right\vert ^{2}=\left\vert e^{-i\delta
		_{^{3}S_{1}(312)}}I_{1}+e^{-i\delta _{^{3}S_{1}(962)}}I_{2}\right\vert
	^{2}=\left\vert I_{1}\right\vert ^{2}+\left\vert I_{2}\right\vert ^{2}+2\cos
	\left( \delta _{^{3}S_{1}(312)}-\delta _{^{3}S_{1}(962)}\right) I_{1}I_{2}.
	\label{Interference}
\end{equation}%

While constructing the radial matrix element in Eq. (\ref{CrossSec}) squared by modulus, we pointed in Eq. (\ref{Interference}) in explicit form the part of scattering $S-$matrix important for analyzing the interference effects (see, for example, Ref. \cite{Rolfs1973}). Note that in the case of non-interfering amplitudes, phase-shift factor $\exp(-\delta_{LJ})$ converts to the unit in the final general expression for the total cross sections \cite{DubBook2019,DubBook2015,PRC2022DTKB}. Below, we demonstrate that the interference term $2\cos\left( \delta _{^{3}S_{1}(312)}-\delta _{^{3}S_{1}(962)}\right) I_{1}I_{2}$  plays an important role in the whole treated energy region, especially for the reaction rate at low temperatures (see insert in Fig. \ref{Fig3}).

In Eq. (\ref{Interference}) the overlapping integral between the initial $%
\chi _{i}$ and final $\chi _{f}$\ states radial functions $I(k) =  \int_{0}^{\infty} \chi^*_ir\chi_f dr $ includes both the interior and asymptotic regions via the continuous functions. The specific behavior of the proton relative motion in the field of the $^{15}$N nucleus is taken into account via the nuclear interaction potential $V(r)$. That is according to the declared above Ansatz. Note that in the cluster model's single-channel approach, the asymptotic behavior of radial WF is actual for their proper normalizing \cite{Rolfs1973}. 

As it is follows from Eq. (\ref{Interference}) in the $E1$ resonance $%
\rightarrow $ ground state transitions\ the interference of $^{3}S_{1}(312)$
and $^{3}S_{1}(962)$\ resonances gives the contribution into the cross
section. The interference is determined by the difference of the $\delta
_{^{3}S_{1}(312)}$ and $\delta _{^{3}S_{1}(962)}$\ phase shifts via the
factor $\cos \left( \delta _{^{3}S_{1}(312)}-\delta _{^{3}S_{1}(962)}\right)
$. We depict the behavior of this factor as a function of energy in Fig. \ref%
{Interfer}$a$ using the phase shifts shown in Fig. \ref{Fig1}. One can
conclude that the contribution of the interfering term into the $E1$
transitions cross section is very significant at the energies up to 2.5 MeV.

\begin{figure}[h]
	\centering
	\includegraphics[width=8.0cm]{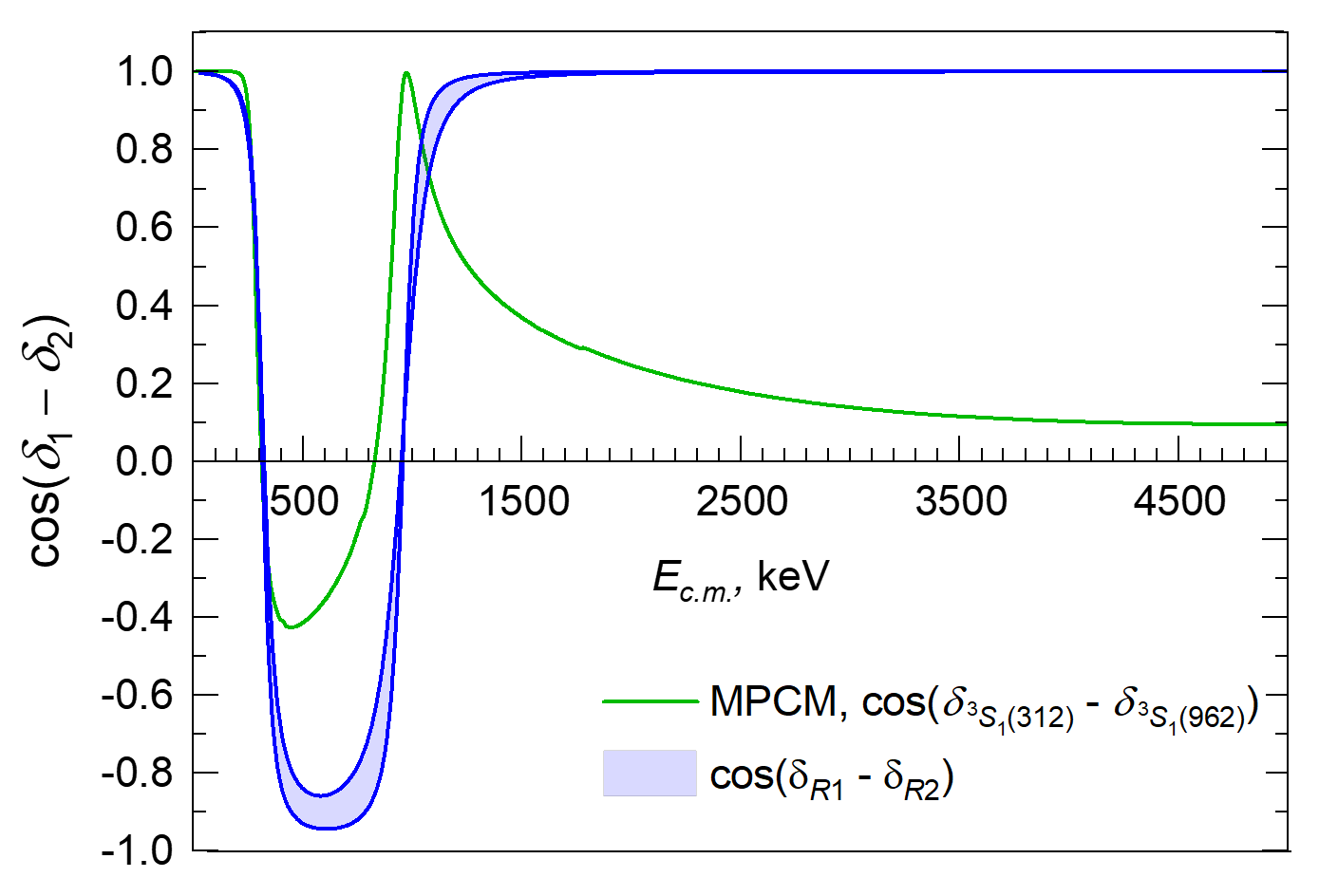} \centering
	\caption{The energy dependence of the factors $\cos \left( \protect\delta %
		_{^{3}S_{1}(312)}-\protect\delta _{^{3}S_{1}(962)}\right) $ (green curve)
		and $\cos \left( \protect\delta _{R1}-\protect\delta _{R2}\right) $ (shaded
		area), respectively. The upper and lower curves of the shaded area are
		obtained with the $\protect\delta _{R1}$ and $\protect\delta _{R2}$ resonant
		phase shifts that are calculated for the experimental $\Gamma _{R1}=91$ keV,
		$\Gamma _{R2}=130$ keV and arbitrary $\Gamma _{R1}=120$ keV, $\Gamma
		_{R2}=250$ keV widths, respectively.}
	\label{Interfer}
\end{figure}

To illustrate the role of the factor $\cos \left( \delta
_{^{3}S_{1}(312)}-\delta _{^{3}S_{1}(962)}\right) $, we replace the $\delta
_{^{3}S_{1}(312)}$ and $\delta _{^{3}S_{1}(962)}$ phase shifts by the
resonant $\delta _{R1}$ and $\delta _{R2}$ (\ref{DeltaR}). In this case, the
factor $\cos \left( \delta _{R1}-\delta _{R2}\right) $ depends on the widths
of the resonances. We calculate the energy dependence of the $\cos \left(
\delta _{R1}-\delta _{R2}\right) $ by varying the width of the resonances
from the experimental $\Gamma _{R1}=91$ keV and $\Gamma _{R2}=130$ keV \cite%
{Ajzenberg1993} up to arbitrary values $\Gamma _{R1}=120$ keV and $\Gamma
_{R2}=250$ keV (the upper and lower curves in Fig. \ref{Interfer},
respectively). The results of the calculations are shown in Fig. \ref%
{Interfer} as a shaded area. We can conclude that the factor $\cos \left(
\delta _{R1}-\delta _{R2}\right) $ is sensitive to the width of the
resonances at the energy interval of about $400-800$ keV and significantly
increases the destructive interference term.

\begin{figure}[b]
	\centering
	\includegraphics[width=8.0cm]{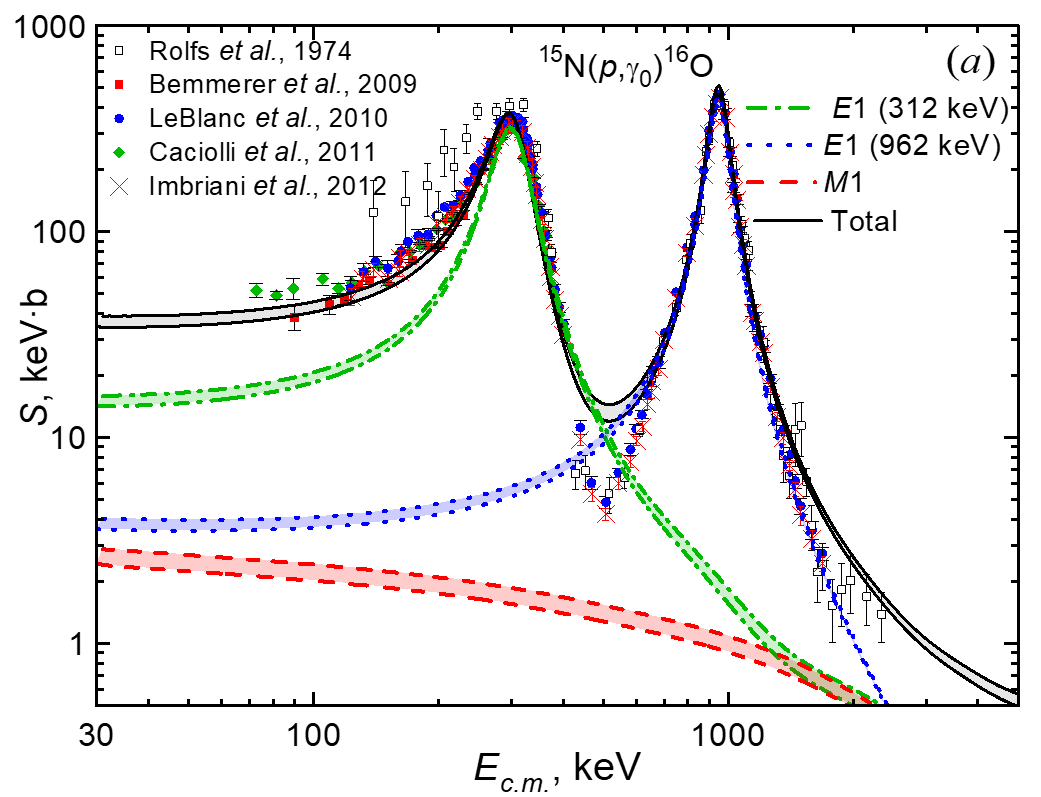} \includegraphics[width=8.0cm]{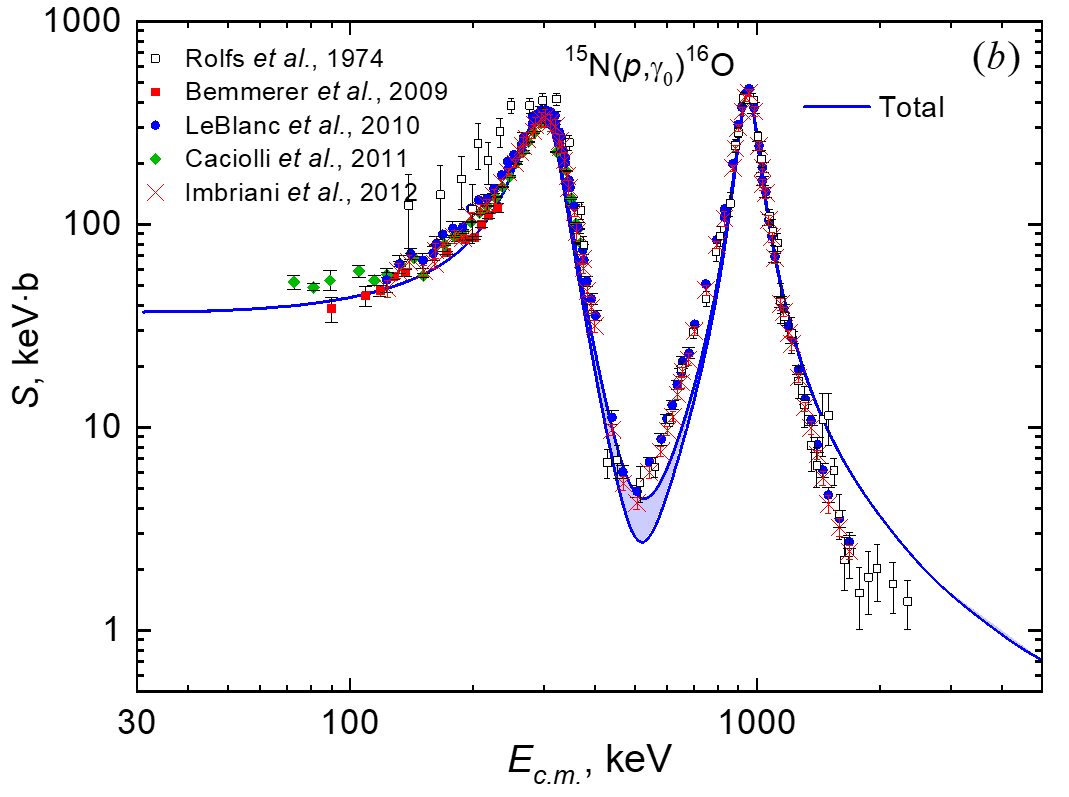}
	\centering
	\caption{(Color online) The astrophysical $S-$factor of radiative $p^{15}$N
		capture on the ground state of $^{16}$O. The shaded areas
		for the partial $E1$, $M1$, and the total $S-$factors correspond to the
		calculations with parameters set I (upper curves) and III (lower curves)
		respectively, from Table \ref{tab:Table1_1}. ($b$) The astrophysical
		$S-$factor is obtained with parameters set II, Table \ref{tab:Table1_1}.
		The shaded area refers to the variation of the cos($\protect\delta _{R1}-$ $%
		\protect\delta _{R2}$) factor for the interference term and corresponds to
		the shaded area in Fig. \ref{Interfer}.}
	\label{Fig2}
\end{figure}

\subsection{Analysis of $S-$factor}

Results of calculations for the astrophysical $S-$factor based on the
potential parameters given in Table \ref{tab:Table1_1} along with the
compilation of experimental data \cite%
{Rolf1974,LeBlanc2010,Imbriani2012,Caciolli2011,Xu2013} are presented in
Fig. \ref{Fig2}$a$. Notice that\ the contribution of 2$^{+}$ level at the
excitation energy 12.97 MeV was considered in Ref. \cite{deBoer2013} using
the $R-$matrix approach. The contribution of this transition to the GS is
much smaller than non-resonance $M1$ $^{3}P_{1}\longrightarrow $ $^{3}P_{0}$
transition. The contribution of the interference of $^{3}S_{1}(312)$ and $%
^{3}S_{1}(962)$\ resonances leads to the significant increase of $S-$factor
at the energies up to 300 keV. One can see the discrepancies between the
experimental data and theoretical calculation at energies where the minimum
of the $S-$factor is observed. This is related to the destructive
interference of $^{3}S_{1}(312)$ and $^{3}S_{1}(962)$\ resonances at this
energy due to the factor $\cos \left( \delta _{^{3}S_{1}(312)}-\delta
_{^{3}S_{1}(962)}\right) $. This factor has a minimum at about 500 keV as it
is depicted in Fig. \ref{Interfer}. The minimum of the $S-$factor is
reproduced precisely within the $R-$matrix approach \cite{deBoer2013} when
the authors are considering the different reaction components' contributions
in the fitting of the $^{15}$N($p,\gamma $)$^{16}$O reaction data (see Fig.
49, \cite{deBoer2013}). Only by using the set of fitting parameters can be
described the region of 0.5 MeV between resonances for the $S-$factor \cite%
{deBoer2013}.

The shaded areas in Fig. \ref{Fig2}$a$ show the range of $S(E)$ changes for
different values of the AC. Thus, the values of transition amplitudes are
governed by the AC. At an energy of 30--60 keV, the $S-$factor is
practically constant 
and the corresponding value can be considered as the $S-$factor at zero
energy. Thus, the theoretical calculation predicts very smooth behavior of $%
S(E)$ at very low energies that converges to $S(0)=35.2(5)$ keV$\cdot $b for
$C_{W}=1.8$ (ANC of $12.85$ fm$^{-1/2}$). The increase of the AC leads to
the increase of $S(0)$. The variation of the AC within the experimental
uncertainties leads to the increase of the $S-$factor up to $S(0)=39.6(8)$
keV$\cdot $b for $C_{W}=2.05$ (ANC of $14.49$ fm$^{-1/2}$). Therefore,
depending on the value of the AC, $S(0)$ varies in the range of $34.7-40.4$
keV$\cdot $b. Our predictions overlap with the value of the $S(0)$ factor
reported in Ref. \cite{Mukhamedzhanov2008}. Note that in Ref. \cite{Son2022}
the value of 29.8(1.1) keV$\cdot $b was obtained in the framework of the
effective field theory. However, \cite{Son2022} describes the experimental $%
S-$factor with the resonances energies $\approx $ 360--370 keV and widths $%
\approx $ 250--350 keV for the first resonance and the resonances energies $%
\approx $ 960--970 keV and widths $\approx $ 155--160 keV for the second
resonance, respectively, which are very far off from the experimental data.

Let us consider, as an example, a different calculation scenario
for the $S-$factor to understand the discrepancies between the experimental
data and theoretical calculation at energies where the $S-$factor has the
minimum. The value and position of the minimum are determined by the
destructive interference of $^{3}S_{1}(312)$ and $^{3}S_{1}(962)$
resonances. We calculate $S-$factor using Eq. (\ref{Interference}) and the
parameters set II from Table \ref{tab:Table1_1}, but replace the factor $%
\cos \left( \delta _{^{3}S_{1}(312)}-\delta _{^{3}S_{1}(962)}\right) $ by $%
\cos \left( \delta _{R1}-\delta _{R2}\right) $. The factor $\cos \left(
\delta _{R1}-\delta _{R2}\right) $ is considered using the same parameters
for the widths as shown in Fig. \ref{Interfer}. The result of contributions
of all transitions to the $S(E)$ is shown by the shaded area in Fig. \ref%
{Fig2}$b$. Thus, we demonstrate
that by varying the factor $-1<\cos \left( \delta _{R1}-\delta _{R2}\right)
<1$ and considering the widths of resonances as parameters, the MPCM can
reproduce the position and value of the $S(E)$ minimum.
It is important mentioning that at very low energies both factors $\cos
\left( \delta _{^{3}S_{1}(312)}-\delta _{^{3}S_{1}(962)}\right) $ and $\cos
\left( \delta _{R1}-\delta _{R2}\right) $ coincide: $\cos \left( \delta
_{^{3}S_{1}(312)}-\delta _{^{3}S_{1}(962)}\right) \simeq \cos \left( \delta
_{R1}-\delta _{R2}\right) \approx 1$ as seen in Fig. \ref{Interfer}. Thus, $%
S(0)$ is not sensitive to this factor, while the variation of this factor is
important for the description of the value and position of the $S(E)$
minimum.

The contribution of the $M1$ non-resonant transition to the GS $S-$factor
comes out of the MPCM through the $^{3}P_{1}$ scattering state and has
significance at the energies {$E_{\text{c.m.}}<500$ keV}. This contribution
increases with the energy decrease. Consideration of the $M1$ transition
requires the experimental $(p,p)$ phase shifts. The elastic {$^{15}$N($p,p$)$%
	^{15}$N} cross sections are measured only at energies higher than 2.7 MeV {%
	\cite{DARDEN1984}}. The low energy elastic scattering experimental data are
desirable to evaluate the intensity of the $^{3}P_{1}\longrightarrow $ $%
^{3}P_{0}$ $M1$ transition.

What is the contribution of each transition to the $S(0)$? $E1$(312 keV)
provides 15.5 keV$\cdot $b (41\%), $E1$(962 keV) gives 3.9 keV$\cdot $b
(10\%), $M1$ does 2.6 keV$\cdot $b (7\%), and the interference term gives
15.7 keV$\cdot $b (42\%).

$R-$matrix calculations reproduce the interference minimum. Our
consideration of the destructive interference of {$^{3}S_{1}$(312)} and {$%
	^{3}S_{1}$(962)} resonances did not reproduce the minimum of the $S-$factor
precisely. The consideration of another state is justified if it has the
same quantum numbers $J^{\pi }=1^{-}$ as the resonances we already
considered. Such a state will interfere with the first two $1^{-}$
resonances. As was already mentioned in Sec II there is such a resonance in
the $^{16}$O spectrum that lays at a very high excitation energy $E_{\text{x}%
}=16.20(90)$ MeV. It is higher than the {$^{3}S_{1}$(312)} resonance by 3.76
MeV and by 3.11 MeV higher than the {$^{3}S_{1}$(962)}. Our numerical
calculations have shown that the effect of this third in the spectrum of $%
1^{-}$ resonances is negligible, and its interference does not influence the
value or position of the $S-$factor minimum.
\begin{table}[h]
	\caption{Experimental data on the astrophysical $S-$factor of the $^{15}$N($%
		p,\protect\gamma $)$^{16}$O reaction. The values of $S(E_{\min })$ listed in
		rows 1 and 2 are taken from Fig. 8 in Ref. \protect\cite{LeBlanc2010}.}
	\label{tab:TableSEx}
	\begin{center}
		\begin{tabular}{ccccc}
			\hline
			& Reference &
			\begin{tabular}{c}
				$E_{\text{c.m.}}$, keV \\
				Experimental range%
			\end{tabular}
			& $E_{\min }$, keV & $S(E_{\min })$, keV$\cdot $b \\ \hline
			1 & Hebbard et al., 1960 \cite{Hebbard} & $206-656$ & 230 & $138.6\pm 15.2$
			\\
			2 & Brochard et al., 1973 \cite{Brochard1973} & $234-1219$ & 256 & $215.1\pm
			27.3$ \\
			3 & Rolfs \& Rodney, 1974 \cite{Rolf1974} & $139-2344$ & 139 & $124.2\pm
			52.6 $ \\
			4 & Bemmerer et al., 2009 \cite{Bemmerer2009} & $90-230$ & 90 & $38.4\pm 5.4$
			\\
			5 & LeBlanc et al., 2010 \cite{LeBlanc2010} & $123-1687$ & 123 & $53\pm 7.1$
			\\
			6 & Caciolli et al., 2011 \cite{Caciolli2011} & $70-370$ & 70 & $52\pm 4$ \\
			7 & Imbriani et al., 2012 \cite{Imbriani2012} & $131-1687$ & 131 & $48.4\pm
			4.8$ \\ \hline
		\end{tabular}%
	\end{center}
\end{table}

In Table \ref{tab:TableSEx} the experimental data for the GS astrophysical $%
S-$factor in the measured energy ranges are given. The experimental range of
energy is dramatically different which leads to the different values of $%
S(E_{\min }).$ In Ref. \cite{Rolf1974} the cross section is measured for the
highest energy. The cross section for the lowest energy $E_{\text{c.m.}}=70$
keV, which is near of the Gamow range, is reported in Ref. \cite%
{Caciolli2011}. It is obvious that extrapolation of the $S(E)$ to the $S(0)$
using each listed experimental energy range will give the different values
of the $S(0)$, sometimes dramatically different.

The determination of $S(0)$ relies on the dual approach of experimental
measurement of the cross section complemented by theoretical interpretation
and extrapolation from the experimental range of energy to the zero energy.
In Table \ref{tab:TableSTher} the estimates of the astrophysical $S-$factor
at zero energy $S(0)$ obtained using the $R-$matrix fits of the different
sets of experimental data, different model calculations, and extrapolation
of the experimental data are listed. By varying the fitting method, authors
obtained different values of $S(0)$, see for example Ref. \cite%
{Mukhamedzhanov2011}. Theoretical evaluation of astrophysical $S(E)$ and its
extrapolation to $S(0)$ are also model dependent, consequently, the
uncertainties in the computed $S-$factor can be significant \cite%
{Yakovlev2010}. The extrapolation is of insufficient accuracy because of the
difficulties in taking full account of the complexities of the reaction
mechanisms \cite{Wiescher2012} as well.
\begin{table}[h]
	\caption{Values of the astrophysical $S(0)$ factor of the $^{15}$N($p,%
		\protect\gamma $)$^{16}$O reaction. The estimations for values of the $S(0)$
		are obtained based of experimental data from references listed in the
		parentheses.}
	\label{tab:TableSTher}
	\begin{center}
		\begin{tabular}{cc}
			\hline
			Reference & $S(0)$, keV$\cdot $b \\ \hline
			Rolfs \& Rodney, 1974 \cite{Rolf1974} &
			\begin{tabular}{c}
				32 (\cite{Hebbard}) \\
				$64\pm 6$ (\cite{Rolf1974})%
			\end{tabular}
			\\
			Barker, 2008 \cite{Barker2008} & $%
			\begin{tabular}{c}
				$\approx $ 50 -- 55 (\cite{Rolf1974}) \\
				$\approx $ 35 (\cite{Hebbard})%
			\end{tabular}%
			$ \\
			Mukhamedzhanov et al., 2008 \cite{Mukhamedzhanov2008} & 36.0 $\pm $ 6 \\
			LeBlanc et al., 2010 \cite{LeBlanc2010} & $39.6\pm 2.6$ \\
			Huang et al.,2010 \ \cite{Huang2010} & 21.1 \\
			Mukhamedzhanov et al., 2011 \cite{Mukhamedzhanov2011} & $33.1-40.1$ \\
			Xu et al., 2013 \cite{Xu2013} & 45$_{-7}^{+9}$ \\
			deBoer et al., 2013 \cite{deBoer2013} & 40 $\pm \ $3 \\
			Dubovichenko et al., 2014 \cite{Dubovichenko2014} & $39.5-43.35$ \\
			Son et al., 2022 \cite{SonNP2022} & 30.4 (\cite{Caciolli2011}) \\
			Son et al., 2022 \cite{Son2022} &
			\begin{tabular}{c}
				$75.3\pm 12.1$ (\cite{Rolf1974}) \\
				$34.1\pm 0.9$ (\cite{LeBlanc2010}) \\
				$29.8\pm 1.1$ (\cite{Imbriani2012})%
			\end{tabular}
			\\
			Present work & $34.7-40.4$ \\ \cline{1-1}
			\multicolumn{2}{c}{Results for $S(0)$ 35.2 $\pm $ 0.5$^{a}$ and 39.6 $\pm $
				0.8$^{b}$ are obtained for AC:} \\
			\multicolumn{2}{c}{$C_{W}$= 1.8$^{a}$ (ANC of 12.85 fm$^{1/2}$) and $C_{W}$=
				2.05$^{b}$ (ANC of 14.49 fm$^{1/2}$).} \\ \hline
		\end{tabular}%
	\end{center}
\end{table}

At ultra-low energies, the energy dependence of the $S-$factor can be
modified by "a screening effect". The Coulomb screening effects in the
laboratory plasma as well as astrophysical environment are discussed in
detail in Refs. \cite{Bertulani2016,Famiano2020,Bertulani2020,Casey2023}.
Despite various theoretical studies conducted over the past two decades, a
theory has not yet been found that can explain the cause of the exceedingly
high values of the screening potential needed to explain the data \cite%
{Spitaleri2016}.

Our expectation is that consideration of the screening will increase $S(0)$. 
The lack of parameters for the screening potential in the $p^{15}$N medium
does not allow us to estimate the role of the screening effect in the $^{15}$%
N($p,\gamma $)$^{16}$O reaction. However, if one considers the estimation of
\cite{Assenbaum1987}, the enhancement of $S-$factor at the energies $\sim $
70 keV corresponding to the LUNA lower data consists of near 11\%.

\section{Reaction Rate}

The reaction rates for nuclear fusion are the critical component for
stellar-burning processes and the study of stellar evolution \cite%
{Wiescher2010}. In stellar interiors, where the interacting particles follow
a Maxwell-Boltzmann distribution, the reaction rate describes the
probability of nuclear interaction between two particles with an
energy-dependent reaction cross section $\sigma (E)$. The reaction rate of
proton capture processes can be written as \cite{Wiescher1999,Iliadis2015}.

\begin{equation}
	N_{A}\left\langle \sigma \nu \right\rangle =N_{A}\left( \frac{8}{\pi \mu }%
	\right) ^{1/2}\left( k_{B}T\right) ^{-3/2}\int \sigma (E)E\exp \left( -\frac{%
		E}{k_{B}T}\right) dE=N_{A}\left( \frac{8}{\pi \mu }\right) ^{1/2}\left(
	k_{B}T\right) ^{-3/2}\int S(E)e^{-2\pi \eta }\exp \left( -\frac{E}{k_{B}T}%
	\right) dE.  \label{Rate1}
\end{equation}%
In Eq. (\ref{Rate1}) $N_{A}$ is the Avogadro number, $\mu $ is the reduced
mass of two interacting particles, $k_{B}$ is the Boltzmann constant, $T$ is
the temperature of the stellar environment, and the factor $e^{-2\pi \eta }$
approximates the permeability of the Coulomb barrier between two point-like
charged particles.

\subsection{$^{15}$N($p,\protect\gamma $)$^{16}$O reaction rate}

\begin{figure}[h]
	\centering
	\includegraphics[width=8.5cm]{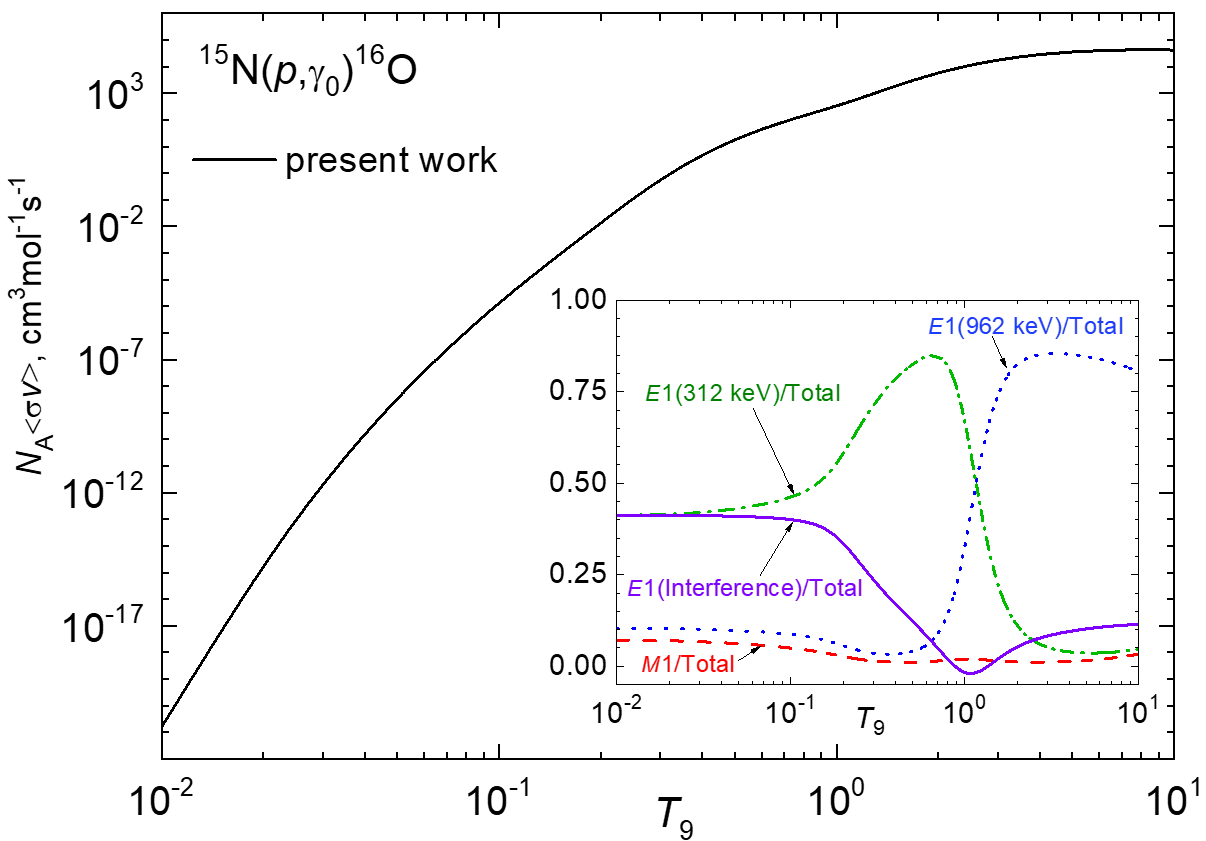} %
	\caption{(Color online) ($a$) The dependence of the reaction rate of the $%
		^{15}$N($p,\protect\gamma _{0}$)$^{16}$O radiative capture on astrophysical
		temperature. The solid curve presents our calculations for the sum of $E1$
		and $M1$ transitions performed for the potentials with the sets of
		parameters from Table \protect\ref{tab:Table1_1}. The inset shows the
		fractional contributions of the reaction rates from the $^{3}S_{1}$
		resonances at 312 keV and 962 keV, respectively, and non-resonance
		transition $^{3}P_{1}\longrightarrow $ $^{3}P_{0}$ with respect to the
		reaction rate of $^{15}$N($p,\protect\gamma _{0}$)$^{16}$O, as a function of
		astrophysical temperature. }
	\label{Fig3}
\end{figure}

The reaction rate can be numerically obtained in the framework of the
standard formalism outline in Ref. \cite{Angulo1999} based on the $S-$factor
that includes the contributions of all transitions shown in Fig. \ref{Fig2}
as well as fractional contributions of $E1$ transitions and $%
^{3}P_{1}\longrightarrow $ $^{3}P_{0}$ $M1$ transition to the $^{15}$N($%
p,\gamma _{0}$)$^{16}$O reaction rate. In Fig. \ref{Fig3} the reaction rate
and fractional contributions of each transition to the reaction rate are
presented. The insert in Fig. \ref{Fig3} shows the contribution of each
resonance and $^{3}P_{1}\longrightarrow $ $^{3}P_{0}$ transition with
respect to the total reaction rate as a function of astrophysical
\begin{figure}[h]
	\centering
	\includegraphics[width=8.0cm]{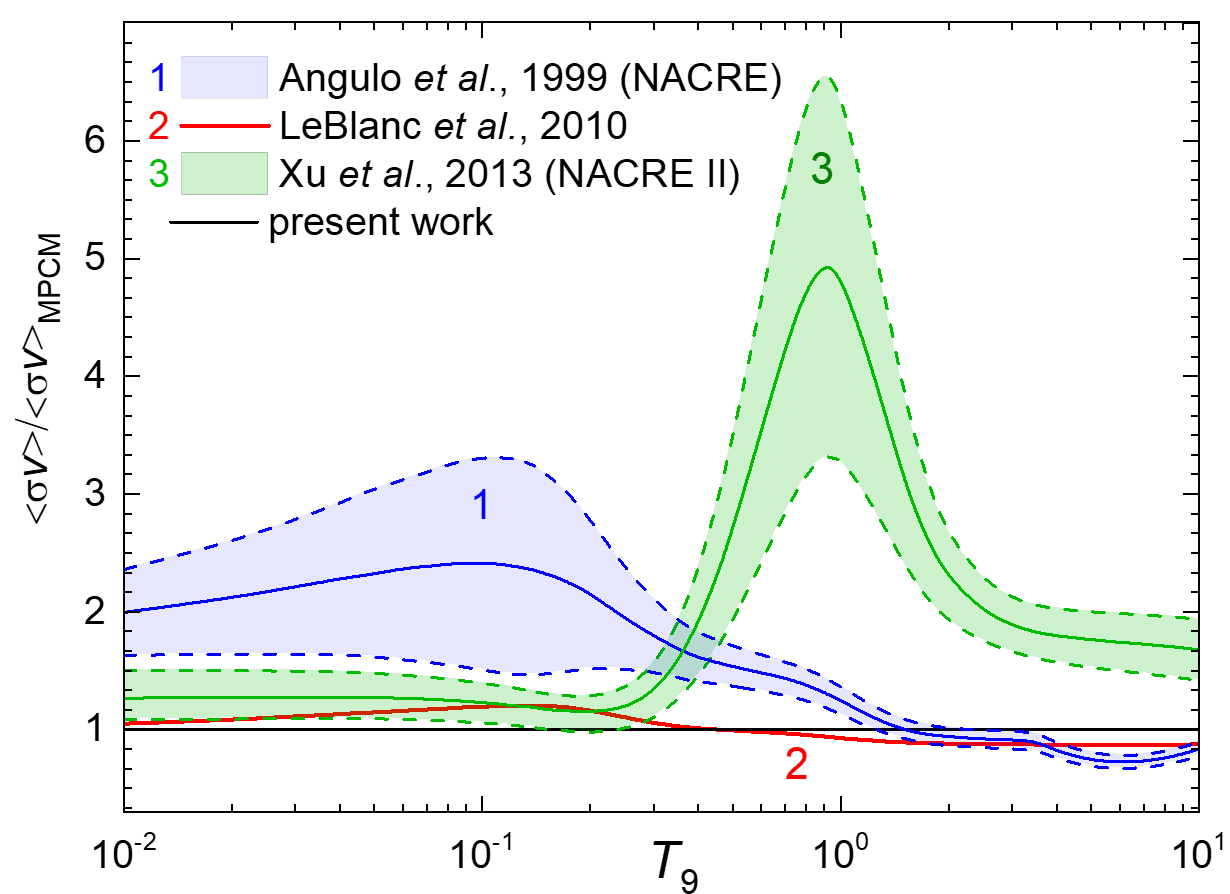}
	\caption{(Color online) The dependence of the ratio of the proton radiative
		capture on $^{15}$N reaction rate from NACRE \protect\cite{Angulo1999}
		(curve 1), \protect\cite{LeBlanc2010} (curve 2), NACRE II \protect\cite%
		{Xu2013} (curve 3) and the present calculation on astrophysical temperature
		in the range of $T_{9}=0.01-10$. The shaded areas within the dashed curves
		represent the uncertainties from NACRE and NACRE II. NACRE \protect\cite%
		{Angulo1999}, LeBlanc et al. \protect\cite{LeBlanc2010}, and present
		calculations are given for the GS transition and NACRE\ II \ parametrization
		includes the GS transition and transitions {via two 2$^{-}$ resonances, and 0%
			$^{-}$ and 3$^{-}$ resonances \protect\cite{Xu2013}.}}
	\label{Fig4}
\end{figure}
\begin{figure}[h]
	\centering
	\includegraphics[width=8.0cm]{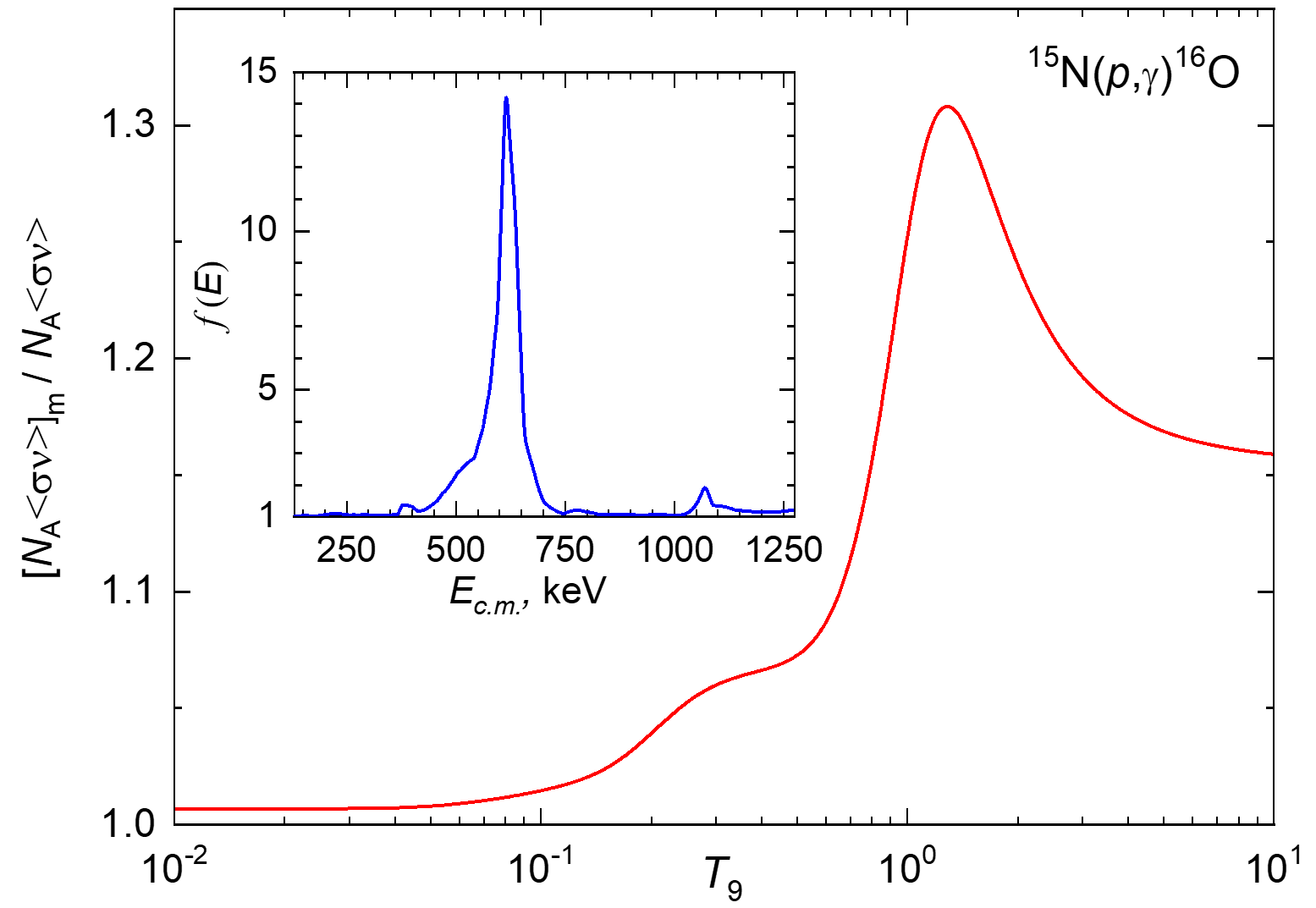} \centering
	\caption{The dependence of the ratio of the total reaction rate which is the
		sum of contributions from the $^{15}$N($p,\protect\gamma _{0}$)$^{16}$O and
		cascade transitions $^{15}$N($p,\protect\gamma _{(6.050)}$)$^{16}$O, $^{15}$%
		N($p,\protect\gamma _{(6.130)}$)$^{16}$O, $^{15}$N($p,\protect\gamma %
		_{(7.117)}$)$^{16}$O and the reaction rate for the GS transition on
		temperature. The experimental data reported in Ref. \protect\cite%
		{Imbriani2012} are used in calculations. The inset shows the dependence of
		the factor $f(E)$ on the proton energy in the c.m.}
	\label{Figratio}
\end{figure}
temperature. Such a presentation is useful to understand the relevance of
each transition at a given temperature. At $T_{9}=0.01$ the fractional
contribution from the 312 keV resonance is 71\%, while the fractional
contributions of the 962 keV resonance and non-resonance transition $%
^{3}P_{1}\longrightarrow $ $^{3}P_{0}$ are 16\% and 13\%, respectively.
However, in contrast, at temperature $T_{9}=10$ the fractional contribution
from the 962 keV resonance is 89\% and contributions of 312 keV resonance
and $^{3}P_{1}\longrightarrow $ $^{3}P_{0}$ transition are commensurate: 6\%
and 5\%, respectively. The $E1$ transitions from 312 keV and 962 keV
resonances have maximal fractional contributions 95\% and 93\% at $T_{9}=0.4$
and $T_{9}=4.1$, respectively. The fractional contribution from
non-resonance transition $^{3}P_{1}\longrightarrow $ $^{3}P_{0}$ increases
with the decrease of the energy.

The interference of the $^{3}S_{1}$(312) and $^{3}S_{1}$(962) resonances
requires special consideration. The solid line in the insert in Fig. \ref%
{Fig3} shows the fractional contribution of the interference term between
the $^{3}S_{1}$(312) and $^{3}S_{1}$(962) resonances to the total reaction
rate. In the Gamow CNO window $T_{9}=0.01-0.03$ the contribution of the
interference term into the total reaction rate is 41\% up to $T_{9}=0.1$. In
the temperature interval $T_{9}=0.01-0.1$ the contribution of the
interference term to the total reaction rate is $\sim $ 40\%, while at $%
T_{9}=3-10$, the contribution of this term does not exceed $\sim $ 11\%. For
the stellar CNO temperature range $T_{9}=0.1-0.5$ the contribution of the
interference term is dropping from $\sim 40$\% to $\sim 12.5$\%. The
destructive interference is observed in the range from $T_{9}=0.89$ to $%
T_{9}=1.36$, but on the level of $\sim $ 2\%. Starting from the end of this
interval up to $T_{9}=10$ a moderate increasing of a constructive
interference is observed from zero up to 11\%.

The dependence of the reaction rate of the $^{15}$N($p,\gamma _{0}$)$^{16}$O
radiative capture as a function of temperature for astrophysical temperature
between $T_{9}=0.01$ and $T_{9}=10$ is shown in Fig. \ref{Fig3}. Results for
the reaction rates for sets I $-$ III are shown by a single solid curve in
Fig. \ref{Fig3}. The reaction rates for $^{15}$N($p,\gamma $)$^{16}$O were
reported earlier in Refs. \cite{Angulo1999,LeBlanc2010,Xu2013}. We
normalized the reaction rate obtained within the $R-$matrix approach \cite%
{LeBlanc2010}, NACRE \cite{Angulo1999}, and NACRE II \cite{Xu2013} reaction
rates by dividing the corresponding data on the reaction rate obtained in
the present calculation. The dependence of these ratios as the function of
astrophysical temperature is shown in Fig. \ref{Fig4}. One can see the
agreement between the reaction rate \cite{LeBlanc2010} obtained for the GS
transition and our calculation. It should be noted the agreement is also
observed for the astrophysical factor: the range of our results for $S(0)$ $%
34.7\leq S(0)\leq 40.4$ keV$\cdot $b and from $37\leq S(0)\leq 42.2$ keV$%
\cdot $b from \cite{LeBlanc2010} are overlapping. In calculations of $S(0)$
in \cite{LeBlanc2010} the used ANC \ of \ $23\pm 3$ fm$^{-1/2}$ is about 3
times larger than the experimental value \cite{Mukhamedzhanov2008}. One can
conclude that the reaction rate is weakly responsive to the value of $S(0)$.

One can see the significant discrepancies in the reaction rates,
particularly, for NACRE \cite{Angulo1999} and NACRE II (includes the GS and {%
	two 2$^{-},$ 0$^{-}$ and 3$^{-}$ resonances}) \cite{Xu2013} data at
temperatures $T_{9}=0.1$ and $T_{9}=1$, respectively, where the ratios reach
the maximums. 
While the difference between \cite{LeBlanc2010} and our $^{15}$N($p,\gamma
_{0}$)$^{16}$O and NACRE II at $T_{9}>0.3$ is understandable. It is puzzling
the significant disagreement in the reaction rates between \cite{LeBlanc2010}
and our, obtained for the GS transition, comparing to NACRE \cite{Angulo1999}
which also parameterized for the GS transition. Maybe this disagreement is
related to the experimental data \cite{Hebbard,Rolf1974} used for the
parametrization of the reaction rate reported in NACRE \cite{Angulo1999}
which were excluded in NACRE II \cite{Xu2013} in favor of the post-NACRE
data.

The reaction rates reported in \cite{Xu2013} include cascade transitions via
two 2$^{-}$ resonances, and 0$^{-}$ and 3$^{-}$ of $^{16}$O resonances in
the $0.40\lesssim E_{R}\lesssim 1.14$ MeV range \cite%
{Hagedorn1957,Gorodetszky1968,Ajzenberg1993,Imbriani2012}. It is stated in
Ref. \cite{Xu2013} the enhancement of the ratio around $T_{9}={1}$ seen in
Fig. \ref{Fig4} is owing to the contribution of the cascade transitions,
which are not included in NACRE \cite{Angulo1999}. In Refs. \cite%
{Imbriani2012,Imbriani2012Err} partial cross sections of the radiative
proton capture to the GS, 1$^{st}(0^{+})$, 2$^{nd}(3^{-}),$ and 4$%
^{th}(1^{-})$ excited states are measured, and converted to the
astrophysical $S-$factors of the $^{15}$N($p,\gamma _{0}$)$^{16}$O, $^{15}$N(%
$p,\gamma _{(6.050)}$)$^{16}$O, $^{15}$N($p,\gamma _{(6.130)}$)$^{16}$O, and
$^{15}$N($p,\gamma _{(7.117)}$)$^{16}$O reactions. The experimental data on (%
$p,\gamma _{1}$), ($p,\gamma _{2}$), ($p,\gamma _{4}$) were reported by
\textit{Imbriani et al.} \cite{Imbriani2012Err} as cascade transitions. We
estimate the contribution of these transitions in the framework of the MPCM
based on \textit{Imbriani's et al}. \cite{Imbriani2012Err} experimental
data, because exact calculations of the cascade transitions are out of the
scope of the present paper.

The experimental data \cite{Imbriani2012,Imbriani2012Err} for $S- $factors
listed in the EXFOR database \cite{JCPRG} (Hokkaido University Nuclear
Reaction Data Centre) were used and interpolated with the Origin Pro 2018
software\ \cite{OriginPro}. The interpolation allows to construct the
energy-dependent factor $f(E)=\frac{\text{Total }S-\text{factor}}{\text{GS }%
	S-\text{factor}}$, where "Total $S-$factor" corresponds to the sum of the $%
^{15}$N($p,\gamma _{0}$)$^{16}$O and the cascade transitions $^{15}$N($%
p,\gamma _{(6.050)}$)$^{16}$O, $^{15}$N($p,\gamma _{(6.130)}$)$^{16}$O, $%
^{15}$N($p,\gamma _{(7.117)}$)$^{16}$O, and "GS $S-$factor" refers to the $%
^{15}$N($p,\gamma _{0}$)$^{16}$O. Introducing this factor into Eq. (\ref%
{Rate1})
\begin{equation}
	\left[ N_{A}\left\langle \sigma \nu \right\rangle \right] _{\text{m}%
	}=N_{A}\left( \frac{8}{\pi \mu }\right) ^{1/2}\left( k_{B}T\right)
	^{-3/2}\int S(E)\times f(E)e^{-2\pi \eta }\exp \left( -\frac{E}{k_{B}T}%
	\right) dE,  \label{Compound}
\end{equation}%
one can estimate the contribution of the cascade transitions. Thus, $\left[
N_{A}\left\langle \sigma \nu \right\rangle \right] _{\text{m}}$ is a
modified GS reaction rate which effectively includes the cascade transitions
to the excited states of $^{16}$O: $^{15}$N($p,\gamma _{(6.050)}$)$^{16}$O, $%
^{15}$N($p,\gamma _{(6.130)}$)$^{16}$O, and $^{15}$N($p,\gamma _{(7.117)}$)$%
^{16}$O.

The dependence of factor $f(E)$ and the ratio of the total experimental $S-$%
factor, that is the sum of the contributions from the GS and the cascade
transitions and the reaction rate for the GS transition on the proton energy
in the c.m. and astrophysical temperature, respectively, is shown in Fig. %
\ref{Figratio}. The factor $f(E)$ increases 14 times at $E_{\text{c.m.}}=600$
keV. At temperatures above $T_{9}=0.3$ the cascade gamma-ray transitions to
the excited bound states contribute to the total reaction rate. Therefore,
the enhancement of the ratio $\left[ N_{A}\left\langle \sigma \nu
\right\rangle \right] _{\text{m}}/N_{A}\left\langle \sigma \nu \right\rangle
_{\text{GS}}$ at $T_{9}>0.3$ is owing to the contribution of the cascade
transitions: $^{15}$N($p,\gamma _{(6.050)}$)$^{16}$O, $^{15}$N($p,\gamma
_{(6.130)}$)$^{16}$O, and $^{15}$N($p,\gamma _{(7.117)}$)$^{16}$O. In the
temperature range of about $T_{9}=1-1.3$ we have a maximum $\sim $\ 30\%
deviation of the total rate from the $^{15}$N($p,\gamma _{0}$)$^{16}$O GS
transition rate and a deviation $\sim $ 15\% at $T_{9}=10$. Thus, our
estimations based on the experimental partial $S-$factors contradict to a
peculiar enhancement behaviour of the ratio reported in NACRE II at $0.4<$ $%
T_{9}<10$. The latter calls for a careful theoretical investigation of
contributions two interfering resonances 2$^{-}$ at 12.530 MeV and 12.9686
MeV and two interfering resonances 3$^{-}$ at 13.142 MeV and 13.265 MeV.

The results of $R-$matrix calculations of the reaction rate \cite%
{LeBlanc2010} was parameterized in the form
\begin{equation}
	N_{A}\left\langle \sigma \nu \right\rangle =\frac{a_{1}10^{9}}{T_{9}^{2/3}}%
	\exp \left[ a_{2}/T_{9}^{1/3}-\left( T_{9}/a_{3}\right) ^{2}\right] \left[
	1.0+a_{4}T_{9}+a_{5}T_{9}^{2}\right] +\frac{a_{6}10^{3}}{T_{9}^{3/2}}\exp
	\left( a_{7}/T_{9}\right) +\frac{a_{8}10^{6}}{T_{9}^{3/2}}\exp \left[
	a_{9}/T_{9}\right]  \label{analytical}
\end{equation}%
and calculations with the parameters from \cite{LeBlanc2010} brought us to $%
\chi ^{2}=20.8$. However, by varying the parameters, we get a much smaller $%
\chi ^{2}=0.4$. The corresponding parameters are given in the first column
in Table \ref{tab:Table2} in Appendix. Parametrization coefficients of the
reaction rate obtained in the framework of MPCM for the analytical
expression (\ref{analytical}) with the parameters from Table \ref%
{tab:Table1_1} are presented in Appendix and are leading to $\chi ^{2}=0.084$%
, $\chi ^{2}=0.086$, and $\chi ^{2}=0.09$ for the sets I, II and III,
respectively. We also present the parametrization coefficients for $\left[
N_{A}\left\langle \sigma \nu \right\rangle \right] _{\text{m}}$ for the set
II.

\subsection{Comparison of rates for proton capture reactions on nitrogen
	isotopes}

There are two stable nitrogen isotopes $^{14}$N and $^{15}$N and all other
radioisotopes are short-lived. Among short-lived isotopes, the longest-lived
are $^{12}$N and $^{13}$N with a half-life of about 11 ms and 9.965 min,
respectively, and they are of nuclear astrophysics interest. A radiative
proton capture on nitrogen isotopes in the reactions $^{12}$N($p,\gamma $)$%
^{13}$O, $^{13}$N($p,\gamma $)$^{14}$O, $^{14}$N($p,\gamma $)$^{15}$O, and $%
^{15}$N($p,\gamma $)$^{16}$O produces the short-lived $^{13}$O, $^{14}$O, $%
^{15}$O isotope with a half-life of $\sim $ 6 ms, $\sim $ 71 s and $\sim $
122 s, respectively, and a stable $^{16}$O nucleus. These radiative capture
reactions caused by the electromagnetic interaction are significantly slower
than reactions induced by the strong interactions. Therefore, these slow
reactions control the rate and time of cycles of oxygen isotopes
nucleosynthesis at particular astrophysical temperatures.

\begin{figure}[h]
	\centering
	\includegraphics[width=8.0cm]{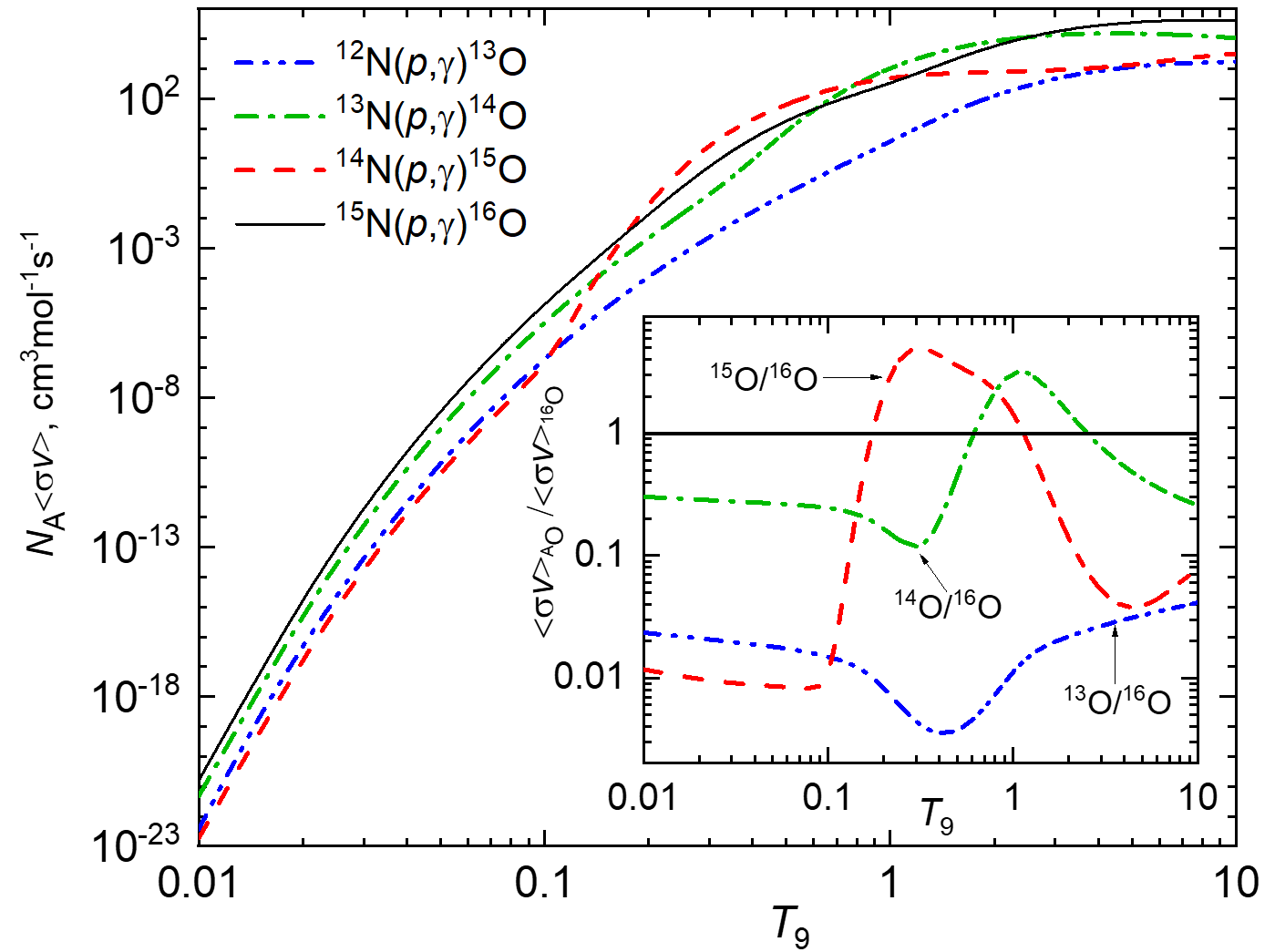}
	\caption{(Color online) The reaction rates of the radiative proton capture
		on nitrogen isotopes leading to the production of oxygen isotopes as a
		function of astrophysical temperature. The insert shows the fractional
		contributions from $^{12}$N($p,\protect\gamma $)$^{13}$O, $^{13}$N($p,%
		\protect\gamma $)$^{14}$O, $^{14}$N($p,\protect\gamma $)$^{15}$O with
		respect to the $^{15}$N($p,\protect\gamma $)$^{16}$O reaction rate as a
		function of astrophysical temperature.}
	\label{Fig5}
\end{figure}

The authors of Ref. \cite{Weischer1989} suggested and discussed three
alternative paths of rapid processes of the CNO cycle leading to the
formation of $^{14}$O through the breakout reactions $^{9}$C($\alpha ,p$)$%
^{12}$N and $^{11}$C($p,\gamma $)$^{12}$N. These three branches of reaction
sequences involve $^{12}$N($p,\gamma $)$^{13}$O, $^{13}$N($p,\gamma $)$^{14}$%
O processes. Thus, these processes are of particular interest for nuclear
astrophysics. In the framework of the MPCM\ the radiative proton capture on
nitrogen isotopes $^{12}$N, $^{13}$N, and $^{14}$N were investigated \cite%
{DubN12,DubN14,PRCDTKB2020}. Because the reactions $^{12}$N($p,\gamma $)$%
^{13}$O, $^{13}$N($p,\gamma $)$^{14}$O, and $^{14}$N($p,\gamma $)$^{15}$O
and the present study of $^{15}$N($p,\gamma $)$^{16}$O are considered on the
same footing within the MPCM, it is useful to compare the reaction rates to
understand the relevance of each process at a given astrophysical
temperature. We compare the reaction rates for$\ ^{12}$N($p,\gamma $)$^{13}$%
O, $^{13}$N($p,\gamma $)$^{14}$O, $^{14}$N($p,\gamma $)$^{15}$O, $^{15} $N($%
p,\gamma $)$^{16}$O reactions involved into the different chains of the CNO
cycles:
\begin{eqnarray}
	&&^{\text{12}}\text{N(}p,\gamma \text{)}^{\text{13}}\text{O -- source of CNO seeds  }\left\{ 
	\begin{tabular}{c}
		\ \ ...$^{11}\text{C}(p,\gamma)$\fbox{$^{12}\text{N}(p,\gamma)^{13}\text{O}$}$(\beta^{+}\nu)^{13}\text{N}$ ...[89\%] or \\ 
		\\ 
		... $^{\text{11}}\text{C(}p,\gamma \text{)}$\fbox{$^{\text{12}}$N$(p,\gamma
			)^{\text{13}}$O}$(\beta ^{+}p\text{)}^{\text{12}}\text{C ... [11\%]}$%
	\end{tabular}%
	\right.   \notag \\
	&&  \notag \\
	&&^{\text{13}}\text{N(}p,\gamma \text{)}^{\text{14}}\text{O -- hot CNO only \
		\ }\ \ \ \ \ \ \ \ \ \ \ \ \ \ \ \ ^{\text{12}}\text{C}(p,\gamma )\fbox{$^{%
			\text{13}}$N$(p,\gamma )^{\text{14}}$O}(\beta ^{+}\nu )\fbox{$^{\text{14}}$N$%
		(p,\gamma )^{\text{15}}$O}(\beta ^{+}\nu )^{\text{15}}\text{N}(p,\alpha )^{%
		\text{12}}\text{C\ }  \notag \\
	&&  \label{Chains} \\
	&&^{\text{14}}\text{N(}p,\gamma \text{)}^{\text{15}}\text{O -- cold and hot
		CNO \ \ \ \ \ }\left\{ \text{\ }%
	\begin{tabular}{c}
		$^{12}\text{C}(p,\gamma )^{13} \text{N}(\beta ^{+}\nu)^{13}\text{C}(p,\gamma)$\fbox{$^{14}\text{N}(p,\gamma )^{15}\text{O}$}$(\beta^{+}\nu)^{15}\text{N}(p,\alpha )^{12}\text{C}$ \\
		\\
		$^{16}\text{O}(p,\gamma )^{17}\text{F}(\beta ^{+}\nu )^{17}\text{O}(p,\alpha)$\fbox{$^{14}\text{N}(p,\gamma )^{15}$O}$(\beta ^{+} \nu)$\fbox{$^{15}\text{N}(p,\gamma )^{16}\text{O}$}
	\end{tabular}%
	\right.   \notag \\
	&&  \notag \\
	&&^{15}\text{N}(p,\gamma)^{16}\text{O -- hot CNO \ \ \
		\ \ \ \ \ \ \ \ \ \ \ \ \ \ \ \ \ \ \ \ \ \ }^{17}\text{O}(p,\gamma)^{18}\text{F}(\beta ^{+}\nu)^{18}\text{O}(p,\gamma )\fbox{$^{15}\text{N}(p,\gamma )^{16}\text{O}$}(\beta^{+}\nu)^{17}\text{F}(\beta^{+}\nu)^{17}\text{O}  \notag
\end{eqnarray}%

The reaction rates of the framed processes are calculated in the framework of the same model, MPCM. 
The temperature windows, prevalence, and significance of each process are considered. The
radiative proton $^{12}$N($p,\gamma $)$^{13}$O, $^{13}$N($p,\gamma $)$^{14}$%
O, $^{14}$N($p,\gamma $)$^{15}$O, $^{15}$N($p,\gamma $)$^{16}$O processes
have the same Coulomb barrier and, as follows from Eq. (\ref{Rate1}), the 
reaction rates will differ only due to the different values of the $S(E)$
and reduced mass $\mu $ of interacting particles in the entrance channel.
The reduced masses of the pairs $p^{12}$N, $p^{13}$N, $p^{14}$N, and $p^{15}$%
N are always less than the proton mass and are within the range $0.9294$ amu 
$\leq \mu \leq 0.9439$ amu. Therefore, the influence of the reduced mass on
the reaction rates of the proton capture on nitrogen isotopes is negligible
and can be omitted. Therefore, the rates of these processes completely
depend on the reaction $S-$factor. Figure \ref{Fig5} gives an overview of
the reaction rates for typical CNO temperature and explosive hydrogen
burning scenarios. The $^{15}$N($p,\gamma $)$^{16}$O reaction is the fastest
one with the biggest rate up to $T_{9}\sim 0.175$ and $p^{14}$N is the
slowest process up to $T_{9}\sim 0.1$ and it controls the rate and time of
nucleosynthesis cycles. One should notice that $^{15}$N($p,\gamma $)$^{16}$O
rate becomes the dominant one at temperature explosive hydrogen burning
scenarios in stars. The analysis of the result presented in the insert in
Fig. \ref{Fig5} leads to the conclusion that only in the temperature windows 
$0.18\lesssim T_{9}\lesssim 1.14$ and $0.66\lesssim T_{9}\lesssim 3$ the
reaction $^{15}$N($p,\gamma $)$^{16}$O is slower than $^{13}$N($p,\gamma $)$%
^{14}$O and $^{14}$N($p,\gamma $)$^{15}$O reactions, respectively. Hence
this slow reaction controls the rate and time of cycles of nucleosynthesis.

It is useful to show the reaction rates of proton radiative capture. The
radiative hydrogen burning induced nucleosynthesis at specific temperatures
has the Gamow peak energy \cite{Fowler1975,Iliadis2015}

\begin{equation}
	E_{0}=\left[ \frac{\pi ^{2}}{\hbar ^{2}}\left( Z_{1}Z_{2}e^{2}\right) ^{2}%
	\frac{\mu }{2}\left( k_{B}T\right) ^{2}\right] ^{\frac{1}{3}}
	\label{GamovE0}
\end{equation}%

which is defined by the condition $\frac{d}{dE}f_{G}(E,T)=0$, where $%
f_{G}(E,T)=e^{-2\pi \eta }\exp \left( -\frac{E}{k_{B}T}\right) $ is a Gamow
function. In the case of the proton and nitrogen isotopes in the entrance
channel $Z_{1}=1$ and $Z_{2}=7$ for (\ref{GamovE0}) in keV for temperature $%
T_{9}$ one obtains

\begin{equation}
	E_{0}=466.4353\left[ \mu T_{9}^{2}\right] ^{\frac{1}{3}},  \label{GamovE}
\end{equation}%

and the effective energy range determined by the Gamow range $\Delta E_{G}$
(in keV) around the Gamow energy $E_{0}$ is

\begin{equation}
	\Delta E_{G}=452.9821\left[ \mu T_{9}^{5}\right] ^{\frac{1}{6}}.
	\label{GamovD}
\end{equation}%

\begin{figure}[h]
	\centering
	\includegraphics[width=7.0cm]{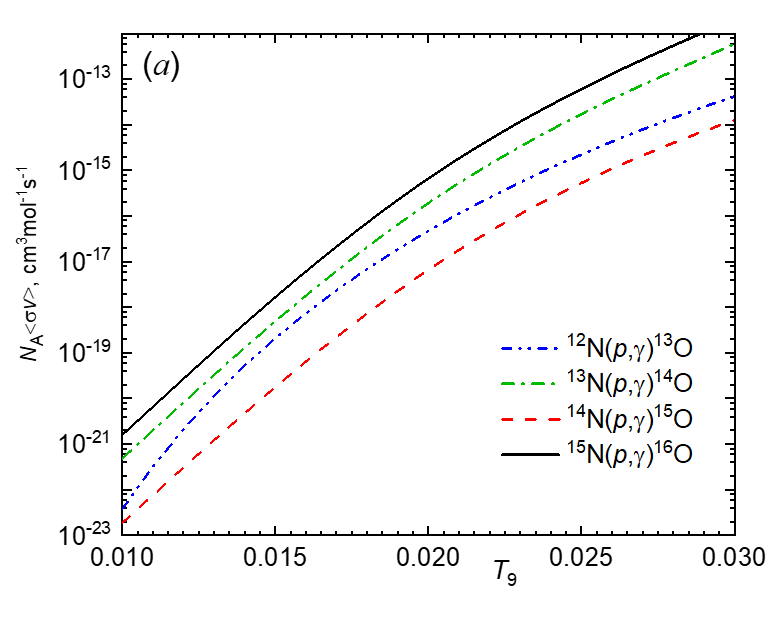} \includegraphics[width=7.0cm]{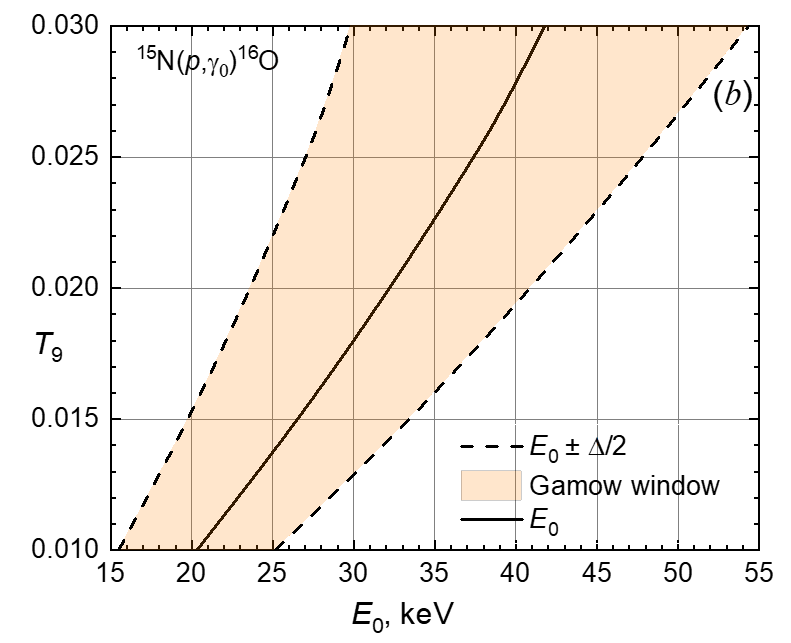}
	\centering
	\caption{(Color online) ($a$) Dependencies of reaction rates of the
		radiative proton capture on nitrogen isotopes on astrophysical temperature
		in the range of $T_{9}=0.01-0.03.$ ($b$)\ The stellar temperatures\ as a
		function of the Gamow energy for CNO cycle $^{12}$N($p,\protect\gamma $)$%
		^{13}$O, $^{13}$N($p,\protect\gamma $)$^{14}$O, $^{14}$N($p,\protect\gamma $)%
		$^{15}$O and $^{15}$N($p,\protect\gamma $)$^{16}$O reactions. }
	\label{Fig6}
\end{figure}

\begin{figure}[h]
	\centering
	\includegraphics[width=7.0cm]{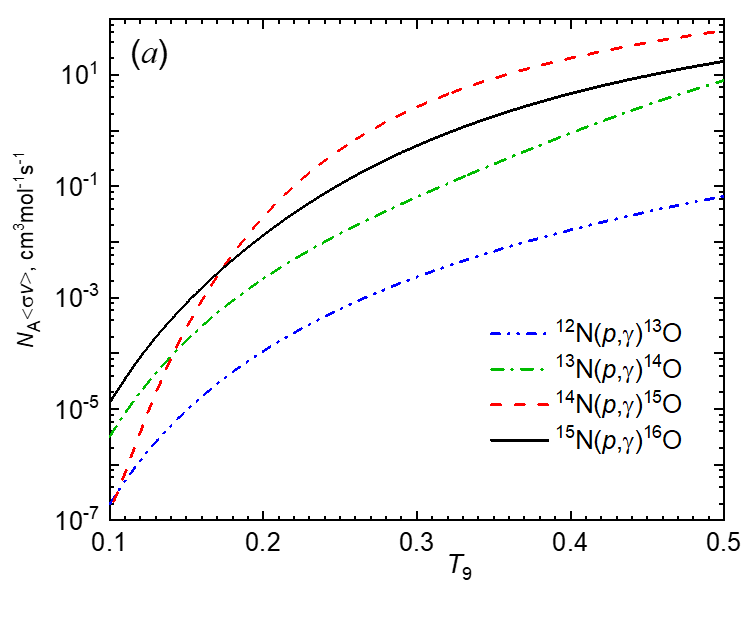} \includegraphics[width=7.0cm]{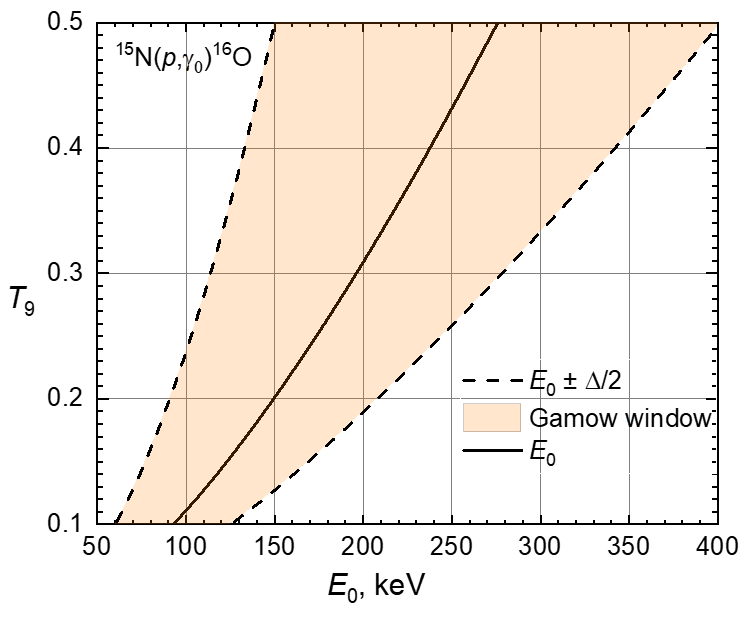}
	\centering
	\caption{(Color online) ($a$) Dependencies of reaction rates of the
		radiative proton capture on nitrogen isotopes on astrophysical temperature
		in the range of 0.1$T_{9}-0.5T_{9}$ ($b$)\ The stellar temperatures\ as a
		function of the Gamow energy for CNO cycle $^{12}$N($p,\protect\gamma $)$%
		^{13}$O, $^{13}$N($p,\protect\gamma $)$^{14}$O, $^{14}$N($p,\protect\gamma $)%
		$^{15}$O and $^{15}$N($p,\protect\gamma $)$^{16}$O reactions. }
	\label{Fig7}
\end{figure}
Thermonuclear reactions occur mainly over the Gamow energy window from $%
E_{0}-\Delta E_{G}/2$ to $E_{0}+\Delta E_{G}/2$ except in the case of narrow
resonances \cite{Iliadis2015}. From Eqs. (\ref{GamovE}) and (\ref{GamovD})
is clear that the Gamow peak energies and ranges for $^{12}$N($p,\gamma $)$%
^{13}$O, $^{13}$N($p,\gamma $)$^{14}$O, $^{14}$N($p,\gamma $)$^{15}$O, $^{15}
$N($p,\gamma $)$^{16}$O reactions are completely determined by the
astrophysical temperature. The variation of the reduced mass within $0.9294$
amu $\leq \mu \leq 0.9439$ amu changes the Gamow peak energy and the energy
range only within 0.5\% and 0.3\%, respectively.

It is useful to present the reaction rate for a particular temperature range
along with the Gamow window of CNO reactions for the radiative proton
capture on nitrogen isotopes. The corresponding results of calculations are
shown in Figs. \ref{Fig6} and \ref{Fig7}. It should be noticed that the main
difficulty in determining reliable reaction rates of $^{12}$N($p,\gamma $)$%
^{13}$O, $^{13}$N($p,\gamma $)$^{14}$O, $^{14}$N($p,\gamma $)$^{15}$O, $%
^{15} $N($p,\gamma $)$^{16}$O reactions for the CNO cycles is the
uncertainty in the very low cross sections at the Gamow range. Developments
within the low-energy underground accelerator facility LUNA in the Gran
Sasso laboratory \cite{Costantini2009} and recent improvements in the
detection setup \cite{Skowronski2023} make taking direct measurements of
nuclear reactions near the Gamow range feasible. This advantage has been
demonstrated in the $^{14}$N($p,\gamma $)$^{15}$O reaction, which was
successfully measured down to energies of 70 keV at LUNA \cite{Lemut2006}.

\section{Conclusion}

We present the results of calculations and analysis of the astrophysical $S-$%
factor and reaction rate for the $^{15}$N($p,\gamma $)$^{16}$O reaction in
the framework of MPCM with forbidden states, including low-lying $^{3}S_{1}$
resonances and $^{3}P_{1}\longrightarrow ^{3}P_{0}$ $M1$ transition. The
intercluster potentials of the bound state, constructed on the basis of
quite obvious requirements for the description of the binding energy and AC
in $p^{15}$N channel of the GS and the scattering potentials describing the
resonances, make it possible to reproduce the behavior of
available experimental data for the total cross section of radiative proton
capture on $^{15}$N nucleus at astrophysical energies. However, it is not so
precise as the $R-$matrix fitting \cite{deBoer2013}.

The interference of $^{3}S_{1}$(312) and $^{3}S_{1}$(962) resonances leads
to the 
significant increase of $S-$factor at the energies up to 300 keV. The
consideration of interfering $^{3}S_{1}$ resonances
and the contribution of $^{3}P_{1}$ scattering wave in $p$ + $^{\text{15}}$N
channel due to $^{3}P_{1}\longrightarrow $ $^{3}P_{0}$ $M1$ transition leads
to an increase of the $S-$factor at low energies. Our result for
the $M1$ transition is related to the corresponding phase shifts. Within our
model we demonstrated that the contribution of the $M1$ transition from the
non-resonance $^{3}P_{1}$ scattering wave 
to the $^{15}$N($p,\gamma _{0}$)$^{16}$O cross section really exists and
provides $\sim $\ 7\% to the $S(0)$. Thus, a systematic and precise
low-energy elastic proton scattering data on $^{15}$N are needed to determine
phase shifts at the energies $E_{c.m.}\lesssim 2$ MeV. The extrapolation of
the $S-$factor at the low energy leads to 35.2 $\pm $ 0.5 keV$\cdot $b and
39.6 $\pm $ 0.8 keV$\cdot $b, depending on the value of the asymptotic
constant, which turned out to be within $34.7-40.4$ keV$\cdot $b. It is
elucidated the important role of the asymptotic constant for the $^{15}$N($%
p,\gamma _{0}$)$^{16}$O process, where the interfering $^{3}S_{1}$(312) and $%
^{3}S_{1}$(962) resonances give the main contribution to the cross section.
A comparison of our calculation for $S(0)$ with existing experimental and
theoretical data shows a reasonable agreement with experimental
measurements. Interestingly, the values of $S(0)$ are consistent with each
other, regardless of utilizing various $R-$matrix method approaches \cite%
{Barker2008,Mukhamedzhanov2008,LeBlanc2010,Mukhamedzhanov2011,Xu2013,deBoer2013}
and present MPCM calculations.
The deviations of $S(0)$ in different approaches are within an accuracy of
the main sources of uncertainties.

The reaction rate is calculated and parameterized by the analytical
expression at temperatures ranging from $T_{9}=0.01$ to $T_{9}=10$ and
compared with the existing rates. The reaction rate has negligible
dependence on the variation of AC, but shows a strong impact of the
interference of $^{3}S_{1}$(312) and $^{3}S_{1}$(962) resonances, especially
at $T_{9}$ referring to the CNO Gamow windows. We estimated the contribution
of the cascade transitions to the reaction rate. The enhancement of the
ratio of the sum of GS and cascade transitions and the GS transition at $%
T_{9}>0.3$ is owing to the contribution of the {$^{15}$N($p,\gamma
	_{(6.050)} $)$^{16}$O, $^{15}$N($p,\gamma _{(6.130)}$)$^{16}$O, and $^{15}$N(%
	$p,\gamma _{(7.117)}$)$^{16}$O }processes.


\bigskip

\textbf{Acknowledgments.} The authors would like to thank A. M. Mukhamedzhanov for useful
discussions.
This research was supported by the Ministry of Science and Higher Education
of the Republic of Kazakhstan under the grant AP09259174.
 \appendix

\section{Parameters for analytical parameterizations}

Parameterization coefficients for the analytical expression (\ref{analytical}%
) of the $^{15}$N($p,\gamma $)$^{16}$O reaction rate date in Ref. \cite%
{LeBlanc2010} and obtained within the framework of MPCM are presented in
Table \ref{tab:Table2}. The sets of parameters I, II, and III from Table \ref%
{tab:Table1_1} lead to the three sets of parametrization coefficients for
Eq. (\ref{analytical}). The Set II$_{\text{m}}$ with $\chi ^{2}=0.093$
presents the parametrization coefficients for Eq. (\ref{analytical}) when
data \cite{Imbriani2012Err} for the transitions $^{15}$N($p,\gamma
_{(6.050)} $)$^{16}$O, $^{15}$N($p,\gamma _{(6.130)}$)$^{16}$O, and $^{15}$N(%
$p,\gamma _{(7.117)}$)$^{16}$O are considered based on Eq. (\ref{Compound}).
\begin{table}[h]
	\caption{Parameters of analytical parametrization of the reaction rate $%
		p^{15}$N capture. The parameters for the reaction rate presented in Ref.
		\protect\cite{LeBlanc2010} with $\protect\chi ^{2}=0.4$ and results obtained
		in the present MPCM calculation.}
	\label{tab:Table2}
	\begin{center}
		\begin{tabular}{cccccc}
			\hline
			& Parameters for \cite{LeBlanc2010} & \multicolumn{1}{|c}{Parameters for MPCM
			} & \multicolumn{2}{|c}{Parameters for MPCM} & \multicolumn{1}{|c}{
				Parameters for MPCM} \\
			& $\chi ^{2}=0.4$ & \multicolumn{1}{|c}{Set I, $\chi ^{2}=0.084$} &
			\multicolumn{1}{|c}{Set II, $\chi ^{2}=0.086$} & Set II$_{\text{m}}$, $\chi
			^{2}=0.093$ & \multicolumn{1}{|c}{Set III, $\chi ^{2}=0.09$} \\ \hline
			$i$ & $a_{i}$ & \multicolumn{1}{|c}{$a_{i}$} & \multicolumn{1}{|c}{$a_{i}$}
			& $a_{i}$ & \multicolumn{1}{|c}{$a_{i}$} \\ \hline
			1 & $0.4874952$ & $1.0375$ & $1.00436$ & $0.90521$ & $0.92832$ \\
			2 & $-15.22289$ & $-15.41934$ & $-15.4231$ & $-15.39278$ & $-15.42226$ \\
			3 & $0.8597972$ & $2.19708$ & $2.17155$ & $2.09762$ & $2.16317$ \\
			4 & $6.734083$ & $0.10981$ & $0.10166$ & $0.47779$ & $0.11569$ \\
			5 & $-2.462556$ & $-0.01995$ & $-0.01655$ & $-0.04484$ & $-0.01654$ \\
			6 & $0.7971639$ & $1.67272$ & $1.72304$ & $1.55827$ & $1.68868$ \\
			7 & $-2.930568$ & $-3.0594$ & $-3.06727$ & $-3.03021$ & $-3.07264$ \\
			8 & $3.224569$ & $4.38681$ & $4.20809$ & $4.93559$ & $3.97887$ \\
			9 & $-11.00680$ & $-12.10183$ & $-12.09517$ & $-12.52017$ & $-12.10662$ \\
			\hline
		\end{tabular}%
	\end{center}
\end{table}

\end{document}